\documentclass[a4paper,11pt]{article}
\usepackage{jheppub}

\usepackage{amsmath,amssymb,bm,dcolumn}
\usepackage{graphicx}

\newcommand{\be}{\begin{equation}}
\newcommand{\ee}{\end{equation}}

\newcommand{\bee}{\begin{eqnarray}}
\newcommand{\eee}{\end{eqnarray}}

\newcommand{\dof}{\mathrm{d.o.f.}}
\newcommand{\feynslash}[1]{/\hspace*{-2mm} #1}
\newcommand{\xoeff}{\tilde X_0}

\def\be{\begin{eqnarray} &&}

\def\ee{\end{eqnarray}}
\def\bew{\begin{widetext}}
\def\ew{\end{widetext}}

\title{On the Quark-Gluon Vertex and Quark-Ghost Kernel: combining Lattice Simulations with Dyson-Schwinger equations}

\author[a]{E.~Rojas}
\author[a]{J.~P.~B.~C.~de Melo}
\author[a]{B.~El-Bennich}
\author[c,d]{O.~Oliveira}
\author[d]{T.~Frederico}

\affiliation[a]{Laboratorio de F\'\i sica Te\'orica e Computacional, Universidade Cruzeiro do Sul, S\~ao Paulo,\\  01506-000  SP, Brazil}
\affiliation[c]{Departamento de F\'{\i}sica, Universidade de Coimbra, 3004-516 Coimbra, Portugal}
\affiliation[d]{Departamento de F\'{\i}sica, Instituto Tecnol\'ogico de Aeron\'autica, DCTA 12.228-900, S\~ao Jos\'e dos Campos, SP, Brazil}

\emailAdd{eduardo.rojas@cruzeirodosul.edu.br}
\emailAdd{joao.mello@cruzeirodosul.edu.br}
\emailAdd{bruno.bennich@cruzeirodosul.edu.br}
\emailAdd{orlando@teor.fis.uc.pt}
\emailAdd{tobias@ita.br}

\abstract{
We investigate the dressed quark-gluon vertex combining two established nonperturbative approaches to QCD: the 
Dyson-Schwinger equation (DSE) for the quark propagator and lattice-regularized simulations for the quark, gluon 
and ghost propagators. The vertex is modeled using a generalized Ball-Chiu ansatz parameterized by a single form 
actor $\xoeff$ which effectively represents the quark-ghost scattering kernel. The solution space of the DSE inversion 
for $\xoeff$ is highly degenerate, which can be dealt with by a numerical regularization scheme. We consider two 
possibilities: (i) linear regularization and (ii) the Maximum Entropy Method. These two numerical approaches yield 
compatible $\xoeff$ functions for the range of momenta where lattice data is available and feature a strong enhancement 
of the generalized Ball-Chiu vertex for momenta below 1~GeV. Our ansatz for the quark-gluon vertex is then used 
to solve the quark Dyson-Schwinger equation which yields a mass function in good agreement with lattice simulations 
and thus provides adequate dynamical chiral symmetry breaking.}

\begin{document}

\maketitle
\flushbottom

\section{Introduction and Motivation}

The strong modification of the dressed gluon and quark propagators from their perturbative form to the infrared (IR) domain ($p^2 \lesssim 1.2$~GeV$^2$)  in 
quantum chromodynamics (QCD) has been the object of intense scrutiny and debate over the past decade. This owes to their intimate connection with dynamical chiral 
symmetry breaking (DCSB) and their eminent role  in confinement scenarios. Consider the dressed gluon two-point function in Landau gauge: many large-volume 
lattice simulations~\cite{Bowman:2004jm,Cucchieri:2007zm,Cucchieri:2010xr,Cucchieri:2011um,Bogolubsky:2009dc,Sternbeck:2012mf,Oliveira:2012eh,Ayala:2012pb}, 
solutions of Dyson-Schwinger equations (DSEs)~\cite{Boucaud:2008ji,Boucaud:2011ug,Aguilar:2006gr,Aguilar:2008xm,Pennington:2011xs,Strauss:2012dg} 
and studies within the refined Gribov-Zwanziger approach~\cite{Dudal:2007cw,Dudal:2008sp,Dudal:2010tf,Cucchieri:2011ig} offer strong evidence for their highly nontrivial 
behavior in the IR, wherefore it is nowadays widely held that the gluon propagator is IR finite. It can be described by a momentum-dependent mass function, $m_g^2(k^2)$, 
whose magnitude in the deep IR is considerable, of the order of $4-16~\Lambda_\mathrm{QCD}^2$, whereas it vanishes as $1/k^2$ for $k^2 \gg \Lambda_\mathrm{QCD}^2$ 
\cite{Oliveira:2010xc}, thereby maintaining full agreement with perturbative QCD. 

Analogously, for the quark propagator such a non-trivial IR behavior has also been evidenced  by the convergence of results on the momentum-dependent wave-function 
renormalization, $Z(p^2)$, and mass function, $M(p^2)$, with numerical solutions of DSEs and simulations of lattice-regularized QCD~\cite{Bowman:2002bm,Bowman:2005vx,Bhagwat:2003vw,Furui:2006ks}.
The origin of the interaction strength at infrared momenta, which guarantees DCSB through the gap equation for the quark, remains unexplained, yet it is well known 
that the support of the DSE kernel must exceed a critical value in order to generate a nontrivial solution~\cite{Atkinson:1986aw,Atkinson:1992tx,Atkinson:1993mz}.

The DSEs are very sensitive to the details of their kernel, one ingredient of which is the nonperturbative quark-gluon vertex. Since the formidable impact
of DCSB on $Z(p^2)$ and $M(p^2)$ is now well established, it is natural to accept that this also be true for the corresponding three-point functions. These functions 
represent the vertices of a given field theory and the impact of DCSB on them was realized early in~\cite{Ball:1980ay}. Ideally, one would solve the DSE for 
the gauge-fermion vertex itself but this proves to be a sisyphian task, as it involves the antifermion-fermion scattering kernel whose skeleton expansion 
contains infinitely many terms; a suitable truncation must be made at an early stage~\cite{Burden:1993gy}. The simplest truncation is known as Rainbow-Ladder 
(RL)~\cite{Roberts:1994dr,Munczek:1994zz,Maris:1999nt,Maris:1997hd}, which satisfies the axial-vector Ward-Green-Takahashi Identity (WGTI)~\cite{Ward:1950xp,
Green:1953te,Takahashi:1957xn} and thus chiral symmetry, whereby the full quark-gluon vertex is replaced by a bare vertex. Given a model for the nonperturbative 
gluon propagator, e.g. Ref.~\cite{Maris:1999nt}, this truncation corresponds to a single gluon exchange re-summed to all orders, hence the coinage `{\em rainbow\/}' 
in the DSE and `{\em ladder\/}' in the Bethe-Salpeter equation (BSE). It has been shown that in the RL truncation, numerical results for physical quantities of light 
pseudoscalar and vector mesons are uniformly over-estimated by $\approx 35$\% in units of mass~\cite{Eichmann:2008ae}.  Steps beyond RL have been taken, 
first neglecting non-Abelian three-gluon interactions~\cite{Bender:1996bb,Bender:2002as,Bhagwat:2003vw,Watson:2004kd,Watson:2004jq,Matevosyan:2006bk} 
which have recently been included~\cite{Fischer:2009jm}.

Practically tractable approaches to the fermion-gauge boson vertex allow for a suitable {\em ansatz\/} that satisfies fundamental symmetries of QCD, amongst 
which chiral symmetry via the axial-vector WGTI. These more general {\em ans\"atze\/} impose constraints of quantum field theory on the vertices, i.e. one insists that 
the vertex must reduce to the bare vertex $\gamma_\mu$ in the large-momentum limit (when dressed propagators can be replaced by bare propagators); it must have 
the same transformation properties as the bare vertex under charge conjugation $C$, parity transformation $P$ and time reversal $T$; it must ensure gauge covariance 
and invariance; and finally one demands that the vertex must be free of kinematic singularities. Moreover, the nonperturbative quark-gluon vertex can always be 
decomposed into longitudinal and transverse components~\cite{Ball:1980ay}, $\Gamma_\mu (p_1, p_2, p_3) = \Gamma^\mathrm{L}_\mu (p_1, p_2, p_3)  + 
\Gamma^\mathrm{T}_\mu (p_1, p_2, p_3)$, [see Eqs.~(\ref{transverse})--(\ref{EQ:Tvertex})]. 

Amongst the vertex models that are largely consistent with the above constraints is the Ball-Chiu vertex~\cite{Ball:1980ay}, which has been successfully employed in 
hadron phenomenology (see Ref.~\cite{Bashir:2012fs} and references therein). Yet, while this vertex satisfies a WGTI as a consequence of gauge invariance, it says 
nothing about the transverse part. In fact, by itself it is insufficient to ensure gauge covariance. It can be augmented following Curtis and Pennington~\cite{Curtis:1990zs}, 
which still does not catch the full extent of the nonperturbative dressing of the quark-gluon vertex; e.g., this extension fails to explain the mass splittings between the $\rho$ 
and $a_1$ parity partners. A minimal extension to reach the mass splitting is to include the dressed-quark anomalous chromomagnetic moment in the vertex~\cite{Chang:2011ei}. 
Moreover, the RL truncation of the DSE/BSE kernels yields numerical results which underestimate the weak decay constant of $D_{(s)}$ and $B_{(s)}$ mesons by 
$25-60$\%~\cite{Nguyen:2010yh}. Numerical improvements in heavy-light $Q \bar q$ systems with merely a Ball-Chiu {\em ansatz\/} are not expected to remedy this 
shortcoming~\cite{Bashir:2012fs,ElBennich:2009vx,ElBennich:2010ha,ElBennich:2012tp}.

Further generalizations of the Curtis-Pennington vertex~\cite{Curtis:1990zs} concentrated on the formulation of the transverse part in Abelian gauge theory 
\cite{Bashir:1994az,Bashir:1995qr,Dong:1994jr} and during the past decade a general nonperturbative construction of the vertex constrained by one-loop perturbation 
theory and multiplicative renormalizability was carried out~\cite{Bashir:1997qt,Bashir:2001vi,Bashir:2011dp}. A promising new unified treatment and solution of the 
longitudinal {\em and\/} transverse WGTI for the Abelian quark-gluon vertex~\cite{Qin:2013mta} confirm these nonperturbative formulations of the vertices guided
by perturbation theory~\cite{Bashir:2011dp}. 

The vertex models just mentioned~\cite{Ball:1980ay,Curtis:1990zs,Bashir:1994az,Bashir:1995qr,Dong:1994jr,Bashir:1997qt,Bashir:2001vi,Bashir:2011dp} are concerned 
with Abelian field theory. They do not account for ghost contributions which enter via the fully-dressed quark-gluon vertex. We recall that the QCD vertex satisfies a Slavnov-Taylor 
identity (STI)~\cite{Slavnov:1972fg,Taylor:1971ff} which is expressed in terms of the dressed quark propagators, $S(p_1)$ and $S(p_2)$, the ghost dressing function, 
$F(p_3^2)$, and the quark-ghost scattering kernel, $H(p_1,p_2,p_3)$, (see, e.g., the discussion in Refs.~\cite{Marciano:1977su,Pascual:1980yu}). 
The latter's Dirac structure is composed of four independent form factors: $X_i(p_1,p_2,p_3)\equiv X_i, i=0,1,2,3$. In the limit within which 
the ghost dressing function is taken to  be unity and all sub-leading contributions in the strong coupling $\alpha_s(k^2)$ are discarded, one essentially recovers 
the Abelian WGTI.

Some of the quark-gluon form factors associated with the longitudinal and transverse components
%
%
have also been obtained in lattice simulations for different kinematic configurations
\cite{Skullerud:2002ge,Skullerud:2003qu,Kizilersu:2006et}. However, the range of space-like momenta on the lattice is so far limited and does not  cover the needs of 
numerical applications involving the quark-gluon vertex. Lacking more insight from lattice-regularized QCD on these form factors, one must resort to symmetries given by 
generalized WGTIs~\cite{Qin:2013mta} and the best possible modeling of the necessary form factors, i.e. the scalar functions, $X_i$, in the quark-ghost scattering 
kernel~\cite{Davydychev:2000rt} and $\tilde{X}_i$, in the matrix-valued scalar amplitudes of the transverse WGTI projections~\cite{Qin:2013mta}. 

Based on 
%
%
%
perturbative one-loop calculations~\cite{Davydychev:2000rt}, Aguilar and Papavassiliou~\cite{Aguilar:2010cn,Aguilar:2013ac} 
calculated the leading nonperturbative corrections to the longitudinal quark-gluon vertex (the ``one-loop dressed" approximation to $X_0$, see Fig.~\ref{fig:quark_ghost}) 
using a particular kinematic configuration. Other configurations are discussed in Ref.~\cite{Cardona2012}. Their approximation leads to a ``ghost-improved" Ball-Chiu 
vertex, which partially ties in the non-Abelian contributions to the dressed quark-gluon vertex. Upon inserting this vertex into the gap equation, no sufficient DCSB is 
found. Nonetheless, in order to establish a workable model for the quark-gluon vertex, the substitution, $F(p_3^2) \to F^2(p_3^2)$, is effectuated in the kernel of the gap 
equation commensurate with an enhancement of the dressed vertex in the IR domain.  

This suggested modification of the kernel 
%
%
%
accounts for two competing effects: {\em i\/}) the approximations made which neglect contributions in the
quark-ghost scattering kernel other than $X_0$. In particular, the {\em ansatz\/} for the leading one-loop dressed expression of $H(p_1,p_2,p_3)$ assumes a simplified 
form of the dressed quark-gluon vertex, namely a vertex $\Gamma_\mu^\mathrm{L}$ that is ghost-free, $F=H=1$. Furthermore, the ghost-gluon vertex is taken at tree-level, 
which is not warranted~\cite{Dudal:2012zx}, and $H(p_1,p_2,p_3)$ is calculated in the chiral limits, $B(p^2)=0$; 
{\em ii\/}) the transverse piece of the vertex is not taken into account yet has been shown to be crucial for sufficient generation of DCSB~\cite{Bashir:2011dp,Chang:2011ei}.

In view of the scarse information on the quark-gluon vertex from first principle calculations, a possible strategy is to
combine the different non-perturbative techniques to solve QCD in order to improve our knowledge on this fundamental vertex.
Herein, we employ the lattice-QCD data for the dressed-quark functions, $A(p^2)$ and $B(p^2)$~\cite{Bowman:2005vx,Furui:2006ks}, 
as well as the (quenched) gluon and ghost propagators~\cite{Bogolubsky:2009dc,Oliveira:2012eh}, $\Delta(q^2)$ and $F(q^2)$,
and numerically extract a momentum-dependent  effective function $\xoeff (q^2)\sim X_0(q^2)$ from the quark gap equation. 
The quark-gluon vertex built from $X_0(q^2)$ is  enhanced in the infrared region and  recovers the perturbative behavior as one approaches larger momenta. Furthermore,
when re-inserted in the gap equation, this quark-gluon vertex 
provides the required DCSB to reproduce the mass function, $M(p^2)$, obtained with lattice-regularized QCD
but it fails to reproduce the lattice quark wave function for momenta below $\sim 700$ MeV.

We have organized this paper as follows: 
we first review the necessary notation and working definitions in Section~\ref{sec1}. In Section~\ref{Sec:LatticeProps},
we present the numerical state-of-the art results for dressed quark, gluon and ghost propagators from lattice-QCD whose fits serve as input and/or constraint for the 
inversion of the quark's DSE. The technical aspects of two different and independent inversion methods as well as the related anatomy of the DSE kernels are
discussed in Section~\ref{inversion}, where the two numerical approaches are based on {\em i\/}) linear regularization; and {\em ii\/}) the Maximum Entropy Method. In Section~\ref{Sec:Closing}, we solve the DSE for the quark propagator using the new vertex,
obtained with the two numerical procedures, and the lattice gluon and ghost propagator and compared the outcome with the lattice 
quark running mass and quark wave function.
Finally, we close with a comparison of both inversion methods and discuss their strengths and limitations in Section~\ref{conclusion}.

\section{Definitions and Notation \label{sec1}} 
\subsection{Dyson-Schwinger equation}

\begin{figure}[t] 
   \centering
   \includegraphics[scale=0.8]{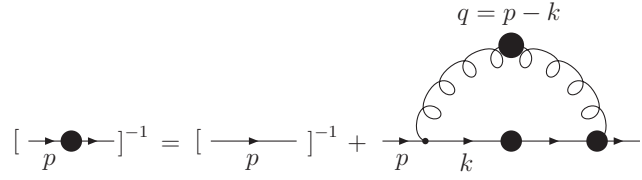} 
   \caption{The Dyson-Schwinger equation for the quark. The solid blobs denote dressed propagators and vertices.}
   \label{FIG:SDE}
\end{figure}

\begin{figure}[t] 
   \centering
   \includegraphics[scale=0.9]{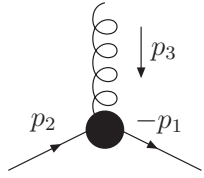} 
    \vspace*{-4mm}  
   \caption{Dressed quark-gluon vertex where the momentum flow indicates that all momenta are incoming.}
   \label{FIG:quark_gluon_vertex}
\end{figure}

Herewith we set the notation used throughout the paper. We begin with the DSE or gap equation in QCD which describes the nonperturbative quark dressing
for a given flavor and is diagrammatically depicted in Fig.~\ref{FIG:SDE}:~\footnote{We employ throughout a Euclidean metric in our notation: 
$\{\gamma_\mu,\gamma_\nu\} = 2\delta_{\mu\nu}$; $\gamma_\mu^\dagger = \gamma_\mu$; $\gamma_5= \gamma_4\gamma_1\gamma_2\gamma_3$, 
tr$[\gamma_4\gamma_\mu\gamma_\nu\gamma_\rho\gamma_\sigma]=-4\, \epsilon_{\mu\nu\rho\sigma}$; $\sigma_{\mu\nu}=(i/2)[\gamma_\mu,\gamma_\nu]$; 
$a \cdot b = \sum_{i=1}^4 a_i b_i$; and $P_\mu$ timelike $\Rightarrow$ $P^2<0$.}
\begin{equation}
S^{-1}(p)  =   \, Z_2 (i\, \feynslash  p + m^{\mathrm{bm}}) + \Sigma (p^2) \ ,
\label{DSEquark}
\end{equation}
where the dressed-quark self-energy contribution is,
\begin{equation}
 \Sigma (p^2) = Z_1\, g^2\! \!\int^\Lambda_k \!\!\ \Delta^{\mu\nu}�(q) \frac{\lambda^a}{2} \gamma_\mu\, S(k) \,\Gamma^a_\nu (-p,k,q) \ .
\end{equation}
The mnemonic shorthand $\int_k^\Lambda\equiv \int^\Lambda d^4k/(2\pi)^4$ represents a Poincar\'e invariant regularization of the integral with the regularization 
mass scale, $\Lambda$, and $Z_{1,2}(\mu,\Lambda )$ are the vertex and quark wave-function renormalization constants.  Owing to the nonperturbative interaction, 
the current-quark bare mass, $m^{\mathrm{bm}}(\Lambda)$, receives corrections from the self-energy $\Sigma (p^2)$, where the integral is over the dressed gluon propagator,  
$\Delta_{\mu\nu}(q)$, the dressed quark-gluon vertex, $\Gamma^a_\nu (-p,k,q)$, and $\lambda^a$ are the usual SU(3) color matrices of the fundamental representation. 
We remind that in Landau gauge the gluon propagator is purely transversal,
\begin{equation}
   \Delta^{ab}_{\mu\nu} (q) =  \delta^{ab} \left( g_{\mu\nu} - \frac{q_\mu q_\nu}{q^2} \right) \Delta ( q^2 ) \ ,
\end{equation}   
and the quark-gluon vertex is given by
\be
  \Gamma^a_\mu (p_1, p_2, p_3) = g \, \frac{\lambda^a}{2} \, \Gamma_\mu (p_1, p_2, p_3) \ ,
\ee
where the momenta, $p_1,p_2$ and $p_3$, are defined in Fig.~\ref{FIG:quark_gluon_vertex} following the convention of Ref.~\cite{Davydychev:2000rt}: $p_1+p_2+p_3 =0$;
$\Gamma_\mu$ represents the Lorentz structure of the vertex. 

With this, the solutions to the gap equation~(\ref{DSEquark}) are of the general form,
\begin{equation}
  \hspace*{-1mm} S(p) = -i\, \feynslash  p \ \sigma_V (p^2) + \mathbb{I}_D \sigma_S(p^2) =  \left [ i\, \feynslash  p\ A(p^2) +\mathbb{I}_D  B(p^2) \right ]^{\!-1}\hspace*{-3mm} ,
  \label{sigmaSV}
\end{equation}
with the renormalization condition,
\be
\left. Z (p^2) = 1/A (p^2)  \right |_{p^2 = \mu^2} = 1 \ , 
\label{EQ:Amu_ren}
\ee 
at large spacelike $\mu^2\gg \Lambda_\mathrm{QCD}^2$. The mass function, 
$M(p^2)=B(p^2\!, \mu^2)/A(p^2\!, \mu^2)$, is independent of the renormalization point $\mu$. In order to make quantitative matching with pQCD, 
another renormalization condition,
\begin{equation}
\left.  S^{-1}(p) \right |_{p^2=\mu^2}  = \  i\  \feynslash  p \ + m(\mu )\, \mathbb{I}_D  \ ,
\label{massmu_ren}
\end{equation}
is imposed, where $m(\mu )$ is the renormalized running quark mass:
\begin{equation}
\label{mzeta} 
  Z_4 (\mu,\Lambda )\, m(\mu)  =  Z_2 (\mu,\Lambda )\, m^{\rm bm} (\Lambda) \ . 
\end{equation}
Herein, $Z_4$ is the renormalization constant associated with the Lagrangian's mass term. In particular, $m(\mu )$ is nothing else but the 
dressed-quark mass function evaluated at one particular deep spacelike point, $p^2=\mu^2$, namely:
\begin{equation}
  m(\mu)  = M(\mu )\,.
\end{equation}
Additionally, one imposes the conditions, 
\bee
\left. \Delta (q^2) \right |_{q^2 = \mu^2} & = & 1 / \mu^2 \ , 
\label{EQ:Glumu_ren}
\ee
on the gluon-dressing function and, 
\bee
\Gamma_\mu (p_1, p_2, p_3)\big |_{p_1^2=p_2^2=p_3^2 = \mu^2} & = & \ \gamma_\mu \ , 
\ee
on the quark-gluon vertex.

%
%
%
%

\subsection{The quark-gluon vertex decomposition}

The matrix-valued vertex function $\Gamma_\mu (p_1, p_2, p_3) = \Gamma^\mathrm{L}_\mu (p_1, p_2, p_3) + \Gamma^\mathrm{T}_\mu (p_1, p_2, p_3)$
can always be decomposed into longitudinal and transverse pieces, the transverse part being defined by
\begin{equation}
  p_3^\mu\  \Gamma^\mathrm{T}_\mu (p_1, p_2, p_3) = 0 \ .
  \label{transverse}
\end{equation}

The full vertex $\Gamma_\mu$ can be written in terms of twelve independent tensors and form factors and we shall follow the decomposition
by Ball and Chiu~\cite{Ball:1980ay}:
\bee
   \Gamma^\mathrm{L}_\mu (p_1, p_2, p_3) & = & \sum^4_{i=1} \lambda_i (p_1, p_2, p_3) \, L^i_\mu (p_1 , p_2) \, ,
   \label{EQ:Lvertex} \\
   \Gamma^\mathrm{T}_\mu (p_1, p_2, p_3) & = & \sum^8_{i=1} \tau_i (p_1, p_2, p_3) \, T^i_\mu (p_1 , p_2) \ .
   \label{EQ:Tvertex} 
\eee
The longitudinal basis can be written as,
\bee
 L^1_\mu (p_1 , p_2) & = &  \gamma_\mu \ , \\
 L^2_\mu (p_1 , p_2) & = &  ( \feynslash p_1 - \feynslash p_2 ) \left( p_1 - p_2 \right)_\mu \ , \\
 L^3_\mu (p_1 , p_2) & = &  i\, ( p_1 - p_2  )_\mu \, \mathbb{I}_D \ ,  \\
 L^4_\mu (p_1 , p_2) & = & \sigma_{\mu\nu} \left( p_1 - p_2 \right)^\nu \ ,
\eee
while the transverse basis is given by,
\bee
 T^1_\mu (p_1 , p_2) & = & i\, \left [ p_{1\mu} \left(p_2 \cdot p_3 \right) -  p_{2\mu} \left(p_1 \cdot p_3 \right) \right ] \mathbb{I}_D  \ , \\
 T^2_\mu (p_1 , p_2) & = &  i\, T^1_\mu  \left( \feynslash p_1 - \feynslash p_2 \right) \ ,  \\
 T^3_\mu (p_1 , p_2) & = & p^2_3 \,  \gamma_\mu - p_{3\mu} \  \feynslash p_3 := p^2_3\,  \gamma_\mu^{\mathrm{T}} \ ,  \\
 T^4_\mu (p_1 , p_2) & = &  - i\, T^1_\mu  (p_1 , p_2) \, \sigma_{\alpha\beta}\, p^\alpha_2 p^\beta_1 \ , \\
 T^5_\mu (p_1 , p_2) & = & \sigma_{\mu\nu} \, p^\nu_3 \ ,  \\
 T^6_\mu (p_1 , p_2) & = & -\gamma_\mu \left( p^2_1 - p^2_2 \right) - \left( p_1 - p_2 \right)_\mu \,  \feynslash p_3 \ , \\
 T^7_\mu (p_1 , p_2) & = & \frac{i}{2} \left( p^2_2 - p^2_1 \right) 
                      \left[ \gamma_\mu \left( \feynslash p_1 - \feynslash p_2 \right) - \left( p_1 - p_2 \right)_\mu \mathbb{I}_D \right]  + 
                      \left( p_1 - p_2 \right)_\mu \sigma_{\alpha\beta} \, p^\alpha_1 \, p^\beta_2 \ , \hspace*{1cm} \\
 T^8_\mu (p_1 , p_2) & = &  i\, \gamma_\mu \, \sigma_{\alpha\beta} \, p^\alpha_1 \, p^\beta_2 - ( p_{1\mu} \, \feynslash  p_2 -  p_{2\mu} \, \feynslash p_1) \ , 
\eee
with $\sigma_{\alpha\beta}  =   \textstyle{\frac{i}{2}}�\left[\, \gamma_\alpha , \gamma_\beta \, \right ]$.
 The quark-gluon vertex satisfies an STI~\cite{Slavnov:1972fg,Taylor:1971ff} to which $\Gamma^\mathrm{T}_\mu$ does not contribute,
\begin{eqnarray}
   p^\mu_3 \ \Gamma_\mu ( p_1, p_2, p_3 ) & = & F(p^2_3) \Big [ �S^{-1} ( -p_1) \, H( p_1, p_2, p_3 ) 
    -    \overline H (p_2, p_1, p_3) S^{-1}(p_2) \Big ] \ ,
\label{STI}
\end{eqnarray}
where the ghost-dressing function $F(q^2)$ is defined by the ghost two-point function,
\begin{equation}
  D^{a b} (q^2) = -\, \delta^{ab} \, \frac{F(q^2)}{q^2} \ , 
\end{equation} 
which is renormalized at $\mu^2 \gg \Lambda_\mathrm{QCD}$, such that
\begin{equation}
  \left. F(q^2) \right |_{q^2 = \mu^2} = 1 \ .
\label{EQ:Fmu_ren}
\end{equation}

The STI relates the quark-gluon vertex with the ghost dressing function, the inverse quark propagator and the quark-ghost
scattering kernel parameterized in terms of the matrix-valued function, $H( p_1, p_2, p_3 )$, and its conjugate, $\overline H ( p_1, p_2, p_3 )$, 
(N.B. see Ref.~\cite{Davydychev:2000rt} for notation and definitions). The decomposition of these two functions in terms of Lorentz covariants 
requires four different form factors,
\begin{eqnarray}
  H( p_1, p_2, p_3 ) & = &  X_0\, \mathbb{I}_D + i\, X_1 \, \feynslash p_1 + i\, X_2 \,\feynslash p_2 + i\, X_3 \, \sigma_{\alpha\beta} p^\alpha_1 p^\beta_2\, ,  \\
 \overline H( p_2, p_1, p_3 ) & = & \overline X_0\, \mathbb{I}_D - i\, \overline X_2\, \feynslash p_1 - i\,\overline X_1\, \feynslash p_2 +
    i\,  \overline X_3 \, \sigma_{\alpha\beta} p^\alpha_1 p^\beta_2\,  ,
 \hspace{6mm}
\end{eqnarray}
where $X_i \equiv X_i ( p_1, p_2, p_3 )$ and $\overline X_i \equiv \overline X_i( p_2, p_1, p_3 )$. 
Perturbative expressions for the form factors $X_i$ have been computed to one-loop order~\cite{Davydychev:2000rt} and yield
$X_0 = 1 + \mathcal{O}(g^2)$ and $X_i = \mathcal{O}(g^2)$, $i = 1, 2, 3$. Thus, $X_0$ is the dominant form factor at large momenta.

The STI, as stated in Eq.~(\ref{STI}), relates the longitudinal form factors $\lambda_i (p_1, p_2, p_3)$ to the quark propagator's 
vector, $A(p^2)$, and scalar, $B(p^2)$, functions~\cite{Aguilar:2010cn} :
\bee
\lambda_1 (p_1, p_2, p_3) & = &
               \frac{F(p^2_3)}{2}� \Big\{�  A(p^2_1) \left[ X_0 + \left( p^2_1 - p_1 \cdot p_2 \right) X_3 \right] 
              +   A(p^2_2) \left[ \overline X_0 + \left( p^2_2 - p_1 \cdot p_2 \right) \overline X_3 \right]  \nonumber \\       
              & + &  B(p^2_1) \left[X_1 + X_2 \right]     +  B(p^2_2) \left[ \overline X_1 + \overline X_2 \right] \Big\} \ ,
              \label{EQ:lambda_1} \\
\lambda_2 (p_1, p_2, p_3) & = &
               \frac{F(p^2_3)}{2\! \left( p^2_2 - p^2_1 \right) } \Big\{� A(p^2_1)\! \left[ \left( p^2_1 + p_1\! \cdot p_2 \right) X_3  - X_0 \right] 
               +   A(p^2_2) \left[ \overline X_0 - \left( p^2_2 + p_1\! \cdot p_2 \right) \overline X_3 \right]  \nonumber \\
               & + & B(p^2_1) \left[X_2 - X_1 \right]  +  B(p^2_2) \left[ \overline X_1 - \overline X_2 \right]   \Big\} \ ,
                \label{EQ:lambda_2} \\
\lambda_3 (p_1, p_2, p_3) & = &
                   \frac{F(p^2_3)}{p^2_1 - p^2_2} \Big\{ A(p^2_1) \left[ p^2_1 \, X_1 + p_1 \cdot p_2 \ X_2\right] -
                     A(p^2_2) \left[ p^2_2\ \overline X_1 + p_1 \cdot p_2 \ \overline X_2 \right]   \nonumber \\
                     & + &  B(p^2_1) \, X_0 -  B(p^2_2) \, \overline X_0  \Big\}  \ ,
                \label{EQ:lambda_3}  \\
\lambda_4 (p_1, p_2, p_3) & = & \frac{- F(p^2_3)}{2}� \Big\{�
                      A(p^2_1) \, X_2 -  A(p^2_2) \, \overline X_2  +  B(p^2_1) \, X_3  - B(p^2_2) \, \overline X_3 \Big\}    \ .                                  
\label{EQ:lambda_4}
\eee

\subsection{The quark-ghost scattering kernel and the Ball-Chiu vertex \label{SubSec:quar_ghost}}

The quark-ghost kernel, i.e. the form factors $X_i$, are only known in pQCD at one-loop order~\cite{Davydychev:2000rt}. 
A nonperturbative model for $H( p_2, p_1, p_3 )$ was proposed~\cite{Aguilar:2010cn} in which only the 
perturbatively dominant form factor, $X_0$, is retained assuming that it is well approximated by its value at a particular 
kinematical point.  This ansatz for $X_0 =\bar X_0 \approx X_0(p^2_3)$ amounts to neglecting the momentum dependence of
$p_1$ and $p_2$ while keeping  the momentum carried by the gluon. In essence, this approximation boils down to simplifying 
the longitudinal vertex of Eqs.~(\ref{EQ:lambda_1})--(\ref{EQ:lambda_4}):
\bee
\lambda_1 (p_1, p_2, p_3)  & = &  \frac{X_0 (p^3_3) \, F(p^2_3)}{2}� \left[ A(p^2_1)  +   A(p^2_2) \right]  \ , \label{bcl1}\\                
\lambda_2 (p_1, p_2, p_3)  & = &  \frac{X_0 (p^3_3) \, F(p^2_3)}{2 \, \left( p^2_2 - p^2_1 \right) } \left[ A(p^2_2)   -  A(p^2_1) \right]  \ , \\
\lambda_3 (p_1, p_2, p_3)  & = &  
             \frac{X_0 (p^3_3) \, F(p^2_3)}{p^2_1 - p^2_2} 
                  \left[ B(p^2_1)  -  B(p^2_2) \right]  \ ,       \\
\lambda_4 (p_1, p_2, p_3) & = &  0 \ . \label{bcl4}
\eee
Note that Eqs.~(\ref{bcl1}) -- (\ref{bcl4}) correspond  to a  Ball-Chiu vertex~\cite{Ball:1980ay} multiplied by the form factor, 
$X_0(p^2_3)$, and the ghost-dressing function $F(p^2_3)$: 
\begin{equation}
\tilde \Gamma^\mathrm{BC}_\mu = X_0 (p^3_3)\,  F(p^2_3)\, \Gamma^\mathrm{BC}_\mu. 
\end{equation}

Turning now to the DSE solutions with the vertex {\em ansatz\/} given by Eqs.~(\ref{bcl1})--(\ref{bcl4}), we find  ($k=p-q,p_1=-p,p_2=k,p_3=q$),
\begin{eqnarray}
  B(p^2) & = & \, Z_2\, m^\mathrm{bm} +\, C_F\, Z_1 g^2\!  \int_q^\Lambda\,  \frac{ \mathcal{K\,}_0 (q)   \mathcal{K\,}_{B} (k,p) } {A^2(k^2) \, k^2 + B^2(k^2)} \ ,
 \label{EQ:SDE_B}  \\
 p^2 A(p^2) & = & \, Z_2 p^2 + C_F\, Z_1 g^2\! \int_q^\Lambda \!  \frac{ \mathcal{K\,}_0 (q)   \mathcal{K\,}_{A} (k,p) }  {A^2(k^2) \, k^2 + B^2(k^2)} \ ,
  \label{EQ:SDE_A} \hspace{1cm}
\end{eqnarray}
where $C_F=4/3$ is the Casimir invariant of the fundamental representation and we have used the following shorthands:
\begin{eqnarray}
\mathcal{K}_0 (q) & = & \Delta (q^2) \, F(q^2) \, X_0 (q^2) \ , 
 \label{K0} \\
\mathcal{K}_A (k,p) & = & \textstyle{\frac{1}{2}} \ A(k^2) \left[\, A(k^2) + A(p^2) \right]  \Big( 3\, k\cdot p - 2\, h(k,p) \Big)  - 2 B(k^2) \, \Delta B( k^2, p^2)\, h(k,p) \nonumber \\ 
  & - &   A(k^2)\,  \Delta A(k^2,p^2)  \left( k^2 + p^2 \right)  h(k,p) \ ,
 \label{EQ:kernel_A} \\
 \mathcal{K}_B (k,p) & = & \textstyle{\frac{3}{2}} B(k^2) \left[ A(k^2) + A(p^2) \right ]  + 2  h(k,p) \Big[ B(k^2) \, \Delta A(k,p)   - A(k^2) \, \Delta B(k,p)  \Big] ,  \hspace{1.3cm} 
 \label{EQ:kernel_B}
\end{eqnarray}
and moreover, 
\bee
 \Delta A(k,p) & = & \frac{A(k^2) - A(p^2)}{k^2 - p^2} \ , \\
 \Delta B(k,p) & = & \frac{B(k^2) - B(p^2)}{k^2 - p^2} \ , \\
 h(k,p) & = & \frac{k^2 p^2 - (k\cdot p)^2}{q^2} \ .
\eee
Note that we have performed a change of variable from the loop-quark momentum, $k$, to the gluon momentum, $q$, in the integrals of 
Eqs.~(\ref{EQ:SDE_B}) and (\ref{EQ:SDE_A}).  These DSEs are our starting point for a numerical extraction of $X_0$ using the lattice-regularized quark, 
gluon and ghost propagators as inputs.  Thus, $X_0$ includes contributions from $X_i$, $i=1,2,3$, as well as from the transverse component of the vertex. 
We shall thenceforth refer to an effective form factor $\xoeff$.

In~\cite{Aguilar:2010cn}  the authors propose a nonperturbative estimation of $X_0$ based on the solution of the integral equation given by the 
`one-loop-dressed'  approximation, see Fig.~\ref{fig:quark_ghost}. Their approximation only takes  into account $X_0 \approx X_0 (p^2_3)$ in Eqs.~(\ref{bcl1})--(\ref{bcl4}) and 
does not provide the IR strength required in the gap equation to generate the expected DCSB. However, a nonperturbative solution of the gap equation compatible 
with DCSB and the asymptotic limit of the mass function, $M(p^2)$, was found {\em if\/} the replacement, $Z_c^{-1}\mathcal{K\,}_0 (q^2) \to F(q^2)\mathcal{K\,}_0(q^2)$, 
or equivalently, $F(q^2) \to F^2(q^2)$, is effectuated in Eqs.~(\ref{EQ:SDE_B}) and (\ref{EQ:SDE_A}), where $Z_c$ is the ghost-function renormalization constant and 
$\mathcal{K\,}_0 (q)$ is defined in Eq.~(\ref{K0}). With this alteration of the gap equation's kernel, a mass-function value of the order $M(0) \approx 300$~MeV 
was obtained but both the pion decay constant, $f_\pi$, and the quark condensate, $\langle \overline q q \rangle$, were 
underestimated by $\sim 20\%$ and $\sim 10 \%$, respectively. 

We mention that, with respect to the approximation just discussed, various kinematic configurations of $H(p_1,p_2,p_3)$ have also been investigated~\cite{Cardona2012,Aguilar:2013ac}.
Of course, the approach to the quark-ghost kernel can be amended in various ways. For example, to compute $X_0$ with the integral equation represented diagramatically 
in Fig.~\ref{fig:quark_ghost} one can replace the tree-level ghost-gluon vertex employed in Ref.~\cite{Aguilar:2010cn}  by the dressed one~\cite{Dudal:2012zx,Boucaud2011,Aguilar:2013March}. 
We have checked this possibility which enhances the quark-gluon vertex but is not sufficient to generate $M(0) \approx 300$~MeV.

\section{The Lattice Propagators \label{Sec:LatticeProps}}   

In this section we discuss the parameterizations of the numerical lattice results for the dressed gluon, ghost and quark propagators
which will be used to solve the DSE inversion. Prior to describing our choice of lattice-propagator data, we would like to remind that 
there exists no complete calculation in the literature using the same set of configurations, i.e. the same set of lattice actions, lattice 
parameters and quark masses.

\subsection{The quenched gluon and ghost propagators \label{SubSec:gluonghostprops}}

For the gluon and ghost propagators we use the lattice data from Ref.~\cite{Bogolubsky:2009dc} generated with $\beta = 5.7$ and 
 the SU(3) Wilson action.  These lattice simulations do not include quark contributions. As discussed in Ref.~\cite{Ayala:2012pb}, taking
into account the fermionic degrees of freedom suppresses the gluon propagator, $\Delta (q^2)$, for momenta $\lesssim 1$~GeV. 
If one takes into account the fermionic degrees of freedom, $\Delta (q^2)$ is suppressed in the IR domain by about  $\sim 20$~\% with
respect  to the quenched result but remains essentially unchanged above 1~GeV.
On the other hand, the dynamical and quenched ghost propagators are about the same within a few percent.
From the point of view of the form factor $\xoeff (q^2)$, replacing the quenched gluon propagator $\Delta (q^2)$ by the dynamical 
gluon propagator $\Delta_\mathrm{dyn.} (q^2)$ in the DSE is equivalent to rescaling the form factor $\xoeff (q^2)$ by the
 momentum-dependent function $\Delta (q^2) / \Delta_\mathrm{dyn.} (q^2)$ which enhances the quark-gluon vertex even further in 
the IR.

Our choice for the quenched data is motivated by the larger physical volume at which the  simulation was performed. Such a large volume 
gives access to smaller momenta in the IR, which it resolves with more details. 
The results in Ref.~\cite{Bogolubsky:2009dc} are also in good agreement with the ones on an enormous lattice volume for the SU(2) gauge
 group~\cite{Cucchieri2007} and corroborated by other large-volume SU(3) simulations, see Ref.~\cite{Oliveira:2012eh} and references therein.

The lattice propagators $\Delta (q^2)$ and $F(q^2)$ were renormalized at $\mu = 4.3$~GeV. In order to solve the DSEs, we use the parameterizations 
discussed in Ref.~\cite{Aguilar:2010cn} for the gluon,
\begin{equation}
   \Delta^{-1}(q^2) = m^2_{gl}(q^2)  + q^2\! \left[ 1 + \frac{13\, C_A g^2_1}{96 \pi^2} \ln \left( \frac{q^2 + \rho_1 m^2_{gl}(q^2)}{\mu^2} \right) 
   \right] \,
   \label{FitGluon}
\end{equation}
with 
\begin{equation}
  m^2_{gl} ( q^2 ) = \frac{m^4}{q^2 + \rho_2 m^2} \ ,
\end{equation}
where $m = 520$ MeV, $\rho_2 = 1.91$, $C_A   = 3$, $g^2_1 = 5.68$, $\rho_1 = 8.55$ and $\mu = 4.3$~GeV, and
\begin{equation}
   F^{-1} (q^2) =  1 + \frac{9\, C_A g^2_2}{192 \pi^2}\ \ln \left( \frac{q^2 + \rho_3 m^2_{gh}(q^2)}{\mu^2}  \right ) 
\label{FitGhost}
\end{equation}
for the ghost-dressing function with 
\begin{equation}
  m^2_{gh} ( q^2 ) = \frac{m^4}{q^2 + \rho_4 m^2} 
\end{equation}
and $g^2_2 = 8.57$, $\rho_3 = 0.25$, $\rho_4 = 0.68$. The quality of the fits can be appreciated in Fig.~5 of Ref.~\cite{Aguilar:2010cn}.

\subsection{The dynamical quark propagator  \label{dynquarkprop}}

Lattice quark propagators in Landau gauge have been studied taking into account light sea-quark degrees of freedom~\cite{Bowman:2005vx,Furui:2006ks}. 
In the following, we only consider the data of Ref.~\cite{Bowman:2005vx}, which corresponds to a lattice simulation with $2+1$ 
flavors of dynamical quarks. The simulation was performed using the improved staggered Asqtad action at $\beta = 7.09$ and, 
of the available  bare-quark mass sets, we choose the one closest to the chiral limit.

\begin{figure}[t]
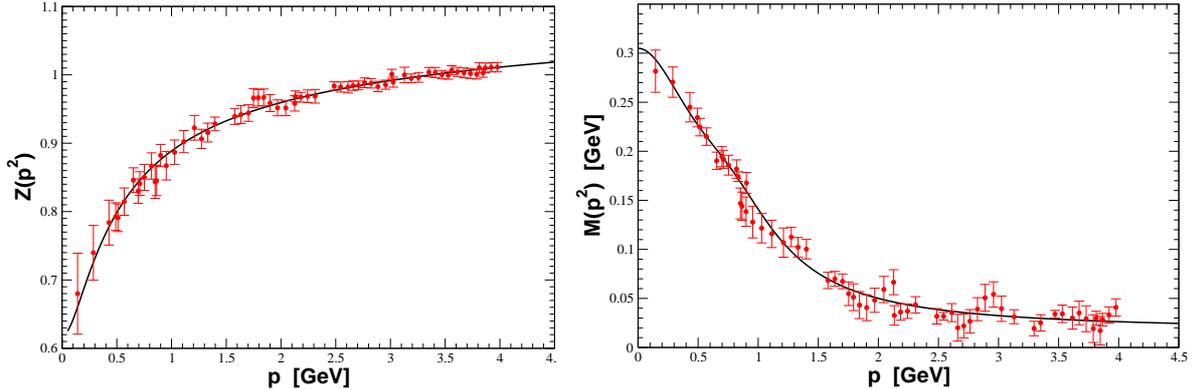
 
\centering
   \hspace*{-6mm}
   \includegraphics[scale=0.3]{z.eps} 
   \includegraphics[scale=0.31]{massa.eps} 
   \caption{Fits to the lattice-regularized QCD data~\cite{Bowman:2005vx}. The left and right plots refer to the quark wave function 
   and mass function, respectively.}
   \label{fig:Z_Mp2}
\end{figure}

A fit of the quark wave function to
\be
  Z(p^2) =  z \left( 1 + \log\left(p^2 + M^2\right) \right)^c 
  \label{FitQuarkWaveFunction}
\ee
yields $\chi^2/\dof =  0.35$ and $z = 0.8576(33)$, $M^2 = 0.395(19)$~GeV$^2$, $c = 0.1235(33)$. The fit of the 
quark mass function to the following parametrization, 
\be 
   M(p^2) =  \frac{ \hat m_q(p^2) }{  \left[ A +  \ln \left( q^2 + \rho\, \hat m^2_q(q^2) \right)  \right]^{\gamma_m} } \ ,
   \label{FitQuarkMass}
\ee
where
\be 
   \hat m_q (p^2) =  \frac{\hat m^3_q}{p^2 + \lambda^2_q}  + m_0  
\ee
and $\gamma_m = 12/ 29$ is the quark anomalous dimension for $N_f = 2$~\footnote{We have fitted the lattice data using $N_f = 3$ and $N_f = 2$.
The parameters as well as the plots are rather similar. However with $N_f = 3$, $M(0)$ is slightly enhanced and the mass functions displays a slight additional 
curvature about $|p| \sim 800$~MeV.}, yields $\chi^2/\dof = 0.86$. The parameter values are $\hat m^3_q = 0.121(11)$~GeV$^3$, 
$\lambda^2_q = 0.311(49)$~GeV$^2$, $m_0 = 0.0343(27)$~GeV, $A = 0.294(80)$ and $\rho = 38.9 \pm 4.8$.  Note that Eq.~(\ref{FitQuarkMass}) is 
essentially the parametrization that describes  the quark mass function from the lattice~\cite{Furui:2006ks} corrected by a logarithmic function to 
reproduce the large-momentum pQCD behavior of $M(p^2)$. The lattice-data points and fits are presented in Fig.~\ref{fig:Z_Mp2}.

The gluon and ghost propagators introduced in Section~\ref{SubSec:gluonghostprops} have been renormalized at $\mu = 4.3$~GeV, whereas the quark 
lattice data is limited to an upper value of $p = 4$~GeV.  In order to be consistent with the renormalization conditions, Eqs.~(\ref{EQ:Amu_ren}), 
(\ref{massmu_ren}), (\ref{EQ:Glumu_ren}) and (\ref{EQ:Fmu_ren}), and to be able to compare our results with Ref.~\cite{Aguilar:2010cn}, we have to 
assume that the fits to $Z(p^2)$ and $M(p^2)$ still describe their correct behavior above 4~GeV. The normalization condition in Eq.~(\ref{EQ:Amu_ren}) 
requires a rescaling of $Z(p^2)$, which implies setting $z = 0.8443$ in Eq. (\ref{FitQuarkWaveFunction}) instead of the previously quoted value.

\section{$\mathbf{\xoeff}$ from Dyson-Schwinger Equations \label{inversion}}

The DSE expressed in Eqs.~(\ref{EQ:SDE_B}) and (\ref{EQ:SDE_A}) is a set of inhomogeneous Fredholm type integral equations of the first kind; see e.g. 
Ref.~\cite{Press:1992zz}. Typically, this class of integral equations does not possess an invertible kernel and Eqs.~(\ref{EQ:SDE_B}) 
and (\ref{EQ:SDE_A}) are no exception. In order to find a solution, i.e. to extract $\xoeff$ from the DSE, the equations have to be regularized. 
We call the reader's attention that this regularization is {\em by no means\/} related to the more fundamental problem of renormalization of 
quantum field theories. In the following Sections, we shall discuss two different and independent inversion methods and their corresponding 
solutions of  the linear system of integral equations, Eqs.~(\ref{EQ:SDE_B}) and (\ref{EQ:SDE_A}), namely: {\em i\/}) linear regularization 
and the {\em ii\/}) Maximum Entropy Method.

For the inversion process, the kernels of the integral equations are built using the fits to the lattice data, i.e. employing Eqs.~(\ref{FitGluon}),  (\ref{FitGhost}),  
(\ref{FitQuarkWaveFunction}) and (\ref{FitQuarkMass}). The measure for the quality of the $\xoeff(q^2)$ solution is the $\chi^2$-comparison of $Z(p^2)$ and 
$M(p^2)$ (obtained from the DSE with exactly this $\xoeff$ reinserted) with the corresponding lattice data values~\cite{Bowman:2005vx}. 
For the range of momenta where the lattice quark propagator is available, i.e. from $\sim 120$~MeV up to $\sim 4$~GeV, the  $\xoeff (p^2)$ 
computed with these two methods are compatible within error estimates. 

The computation of $\xoeff$ requires the definition of  $Z_1$, $Z_2$, the current quark mass and the strong coupling constant. 
Before going into the details of the numerical inversion, let us describe how we determine these numbers. For the strong coupling constant we take  
$\alpha_s ( \mu^2 ) = g^2( \mu^2 ) / 4 \pi = 0.295$~\cite{Boucaud2006,Boucaud2009} at $\mu=4.3$~GeV. Note that this is not the usual value,
$\alpha_s (4.3 \mbox{ GeV} ) = 0.22$, but our choice for the strong coupling constant will allow for a direct comparison with the results of 
Ref.~\cite{Aguilar:2010cn,Aguilar:2013ac}. In what concerns  $\xoeff$, a different 
definition for $\alpha (\mu)$ requires, accordingly, the rescaling of $\xoeff$, i.e. our $\xoeff$ should be multiplied by $1.34$ to convert to the form factor 
at the right scale. On the other hand, the renormalization constant $Z_1$ always appears associated with the form factor $\xoeff$. Thus, in the inversion 
of the DSE we always compute the product, $Z_1 \xoeff$, and therefore do not need to define $Z_1$. Henceforth, whenever we refer to $\xoeff$ 
we imply $Z_1 \xoeff (q^2)$.

The quark wave function renormalization constant $Z_2$ can be computed using the r.h.s. of Eq.~(\ref{EQ:SDE_A}) and requiring the condition 
in Eq.~(\ref{EQ:Amu_ren}) to be satisfied exactly at the renormalization point. We furthermore impose the renormalization condition (\ref{massmu_ren}) 
where we are constrained by the lattice-data points~\cite{Bowman:2005vx}. 
Thus, at $\mu = 4.3$~GeV we set:
\begin{equation}
 m(\mu )  \equiv \frac{B_\mathrm{Latt.}(\mu^2 )}{A_\mathrm{Latt.}(\mu^2 )} = 25.2 \ \mathrm{MeV} \ ;  \ \
  A(\mu^2) \equiv   A_\mathrm{Latt.}( \mu^2)  = 1 .
\label{renormcondition} 
\end{equation}

\subsection{$\mathbf{\xoeff}$ from linear regularized Dyson-Schwinger equations \label{DSEregularized}}

The integrals in Eqs.~(\ref{EQ:SDE_B}) and (\ref{EQ:SDE_A}) can be performed using a Gauss-Legendre quadrature. The integral equations are thus 
transformed into a linear system of (coupled) equations which proves to be ill-defined, as the matrix to be inverted has 
vanishing or very small eigenvalues. However, if the linear system is regularized one can extract $\xoeff$ and, when reinserted in the 
DSE,  check whether its solution is compatible with the quark propagator from the lattice.  In this section, we consider a linear type of regularization. 

Let us for the moment ignore the contribution of the quark mass\footnote{For the computation of $\xoeff$ we use $Z_2 m^{bm} = 14$~MeV. 
Note, however, that the inversion is not sensitive to $7 \lesssim Z_2 m^{bm} \lesssim 20$ MeV.} and write Eq.~(\ref{EQ:SDE_B}) in form of a matrix equation,
\be 
   B =  \overline{\mathcal{K\,}}_B \, \xoeff \ ,
\ee   
where from now on we absorb the factor $C_F g^2 \mathcal{K\,}_0 (q)$ $\left [ A^2(k^2) \, k^2 + B^2(k^2) \right ]^{-1}$ into the kernel, $\overline{\mathcal{K\,}}_B$. 
This equation can be re-expressed as, 
\be
   \xoeff = \frac{B}{\epsilon} + \xoeff - \frac{\overline{\mathcal{K\,}}_B}{\epsilon} \, \xoeff \, ,
\ee   
which defines the following sequence of ``pseudo-solutions" $X^{(i)}_0$ with $i = 0, 1,\dots N$ as
\be
   \xoeff^{(i+1)} = \frac{B}{\epsilon} + \xoeff^{(i)} - \frac{\overline{\mathcal{K\,}}_B}{\epsilon} \,  \xoeff^{(i)} \ ,
   \label{EQ:solve_linear}
\ee

The iterative process is initialized setting $\xoeff^{(0)} = 1$ and stopped whenever $|\xoeff^{(N+1)}\! -\, \xoeff^{(N)}|$ is of the order of the machine precision. 
In the following, we consider a constant regularization parameter~$\epsilon$.

\begin{figure}[t!] 
\vspace*{-5mm}
\centering 
  \includegraphics[scale=0.7]{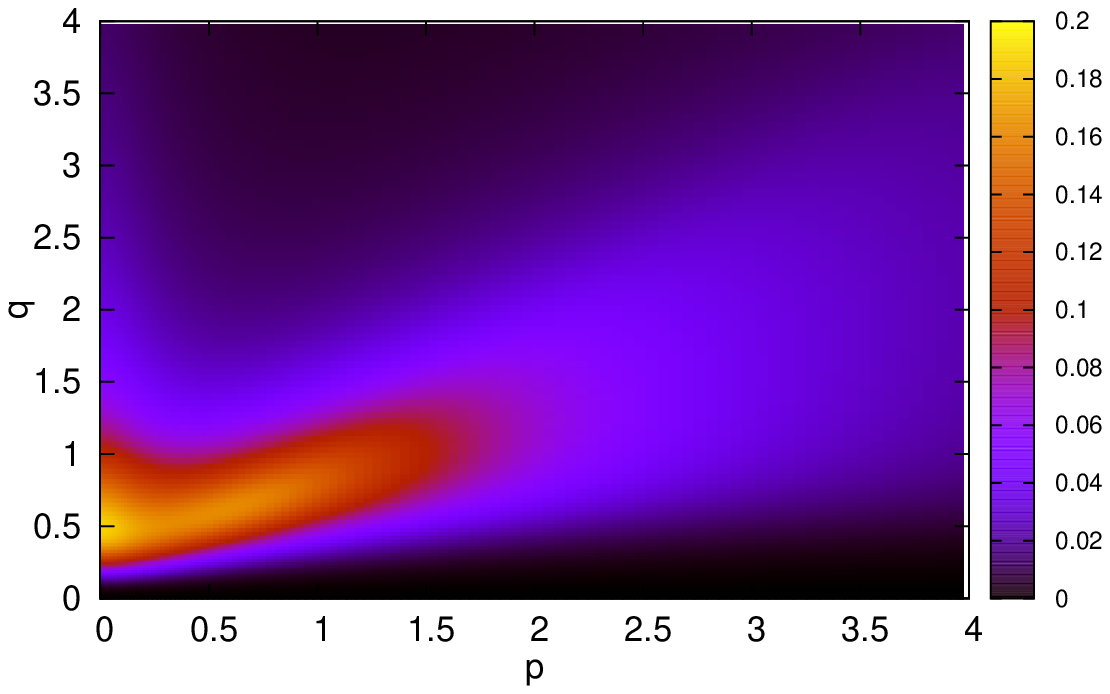}   \\ \vspace*{-1cm}
   \includegraphics[scale=0.7]{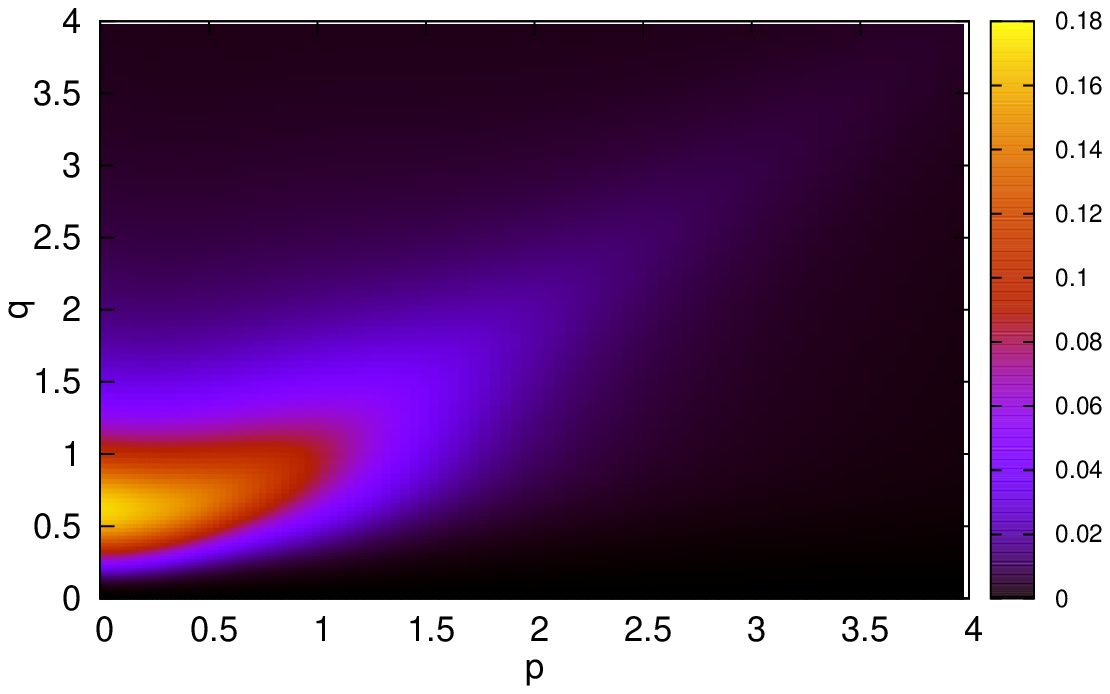}  
   \caption{The kernels $\overline{\mathcal{K\,}}_A (p,q)/p^2$ of Eq.~(\ref{EQ:kernel_A}) (upper graph) and $\overline{\mathcal{K\,}}_B (p,q)$ of Eq.~(\ref{EQ:kernel_B})
     (lower graph) computed with fits to the lattice-quark propagators given by Eqs.~(\ref{FitQuarkWaveFunction}) and (\ref{FitQuarkMass}). }
   \label{fig:kernels}
\end{figure}

The DSE of Eqs.~(\ref{EQ:SDE_B}) and (\ref{EQ:SDE_A}) can be inverted by means of two different regulators, $\epsilon_A$ and $\epsilon_B$. 
However, instead of computing two independent iterations,
\bee 
 \tilde X^{(i)}_{0A}  & \ \longrightarrow & \ \mbox{ from Eq. (\ref{EQ:SDE_A}) } \ , \nonumber \\
 \tilde X^{(i)}_{0B}  & \ \longrightarrow & \ \mbox{ from Eq. (\ref{EQ:SDE_B}) } \ , \nonumber
\eee
we set, at each step, 
\be
  \tilde X^{(i+1)}_0 =  \textstyle{\frac{1}{2}} \left( \tilde X^{(i)}_{0A} + \tilde X^{(i)}_{0B} \right) \ ,
\ee 
to solve simultaneously both equations.  With this superposition, we hope to suppress undesirable effects arising from the kernels' zero modes.

The angular integration was performed with 500~Gauss-Legendre points and we checked the numerical stability  of the  integration. 
For the integration over the momenta, we increased the number of Gauss-Legendre points until reaching a stable answer for $\xoeff$.
We observe no change in the numerical output for a number of integration points above 100. However, all the numerical results 
reported here were computed taking 250 Gauss-Legendre points.

\subsubsection{The kernels of the Dyson-Schwinger equations \label{Sec:kernels_dse}}

The inversion of the DSE does not necessarily provide a reliable evaluation of the form  factor $\xoeff$ over the full momentum range of 
available lattice data. In discretized form, the integrals in Eqs.~(\ref{EQ:SDE_B}) and  (\ref{EQ:SDE_A}) are given by, 
\bee
  B(p^2) & = & Z_2\, m^\mathrm{bm} +  \overline{\mathcal{K\,}}_B (p,q_i)\, \xoeff (q_i) \ , \\
  A(p^2) & = & Z_2 + \overline{\mathcal{K\,}}_A (p,q_i)\, \xoeff (q_i) /p^2  \, , 
\eee  
where a sum $\sum_i$ over $q_i$ is implicit. Note that the kernels, $\overline{\mathcal{K\,}}_A$ and $\overline{\mathcal{K\,}}_B$, now include
the Gauss-Legendre quadrature weights. Computed  using the fits to the lattice-values for $A(p^2)$ and $B(p^2)$, they are graphically displayed 
in Fig.~\ref{fig:kernels} as functions of $p$ and $q$. In fact, Fig.~\ref{fig:kernels} clearly implies that the major contribution of $\xoeff$ to the DSE 
comes from the momentum range, $\sim 0.2~\mathrm{GeV} < q <\ \sim 1.5$~GeV, whereas momenta above 4~GeV and below $\sim 100$~MeV 
are negligible. It follows that the inversion provides some reliable information on the  longitudinal component of the quark-gluon vertex in the 
region $\sim 0.1 -4$~GeV, yet this is not true for $\xoeff$ in the deep IR and UV domains. 

We recall that the lattice data ranges from $141$~MeV up to $4$~GeV for the quark propagator and up to $4.5$~GeV for the gluon and ghost propagators. 
Furthermore, our choice of renormalization point is $\mu = 4.3$. This value of $\mu$ enables a direct comparison of our results with the non-perturbative 
modeling of $X_0 (q^2)$ obtained in Ref.~\cite{Aguilar:2010cn}.

The functional behavior of the kernel also helps us  to define an ultraviolet cutoff, $\Lambda$, in the numerical evaluation of the integrals in
Eqs.~(\ref{EQ:SDE_B}) and (\ref{EQ:SDE_A}). Given that momentum contributions, $q \gtrsim 4$~GeV, are subleading at the most, we consider
inversions with the cutoffs $\Lambda = 5$, $6$ and $7$~GeV.

\subsubsection{The form factor $\mathbf{\xoeff}$}

\begin{figure}[t!] 
\vspace*{-7mm}
   \centering
   \includegraphics[scale=0.75]{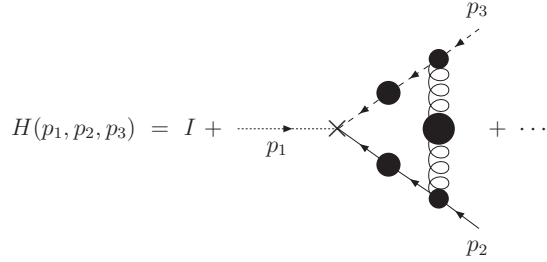} 
    \vspace*{-3mm} 
   \caption{The quark-ghost kernel:  one-loop dressed approximation.}
   \label{fig:quark_ghost}
\end{figure}

\begin{figure}[t!] 
   \centering
   \includegraphics[scale=0.3]{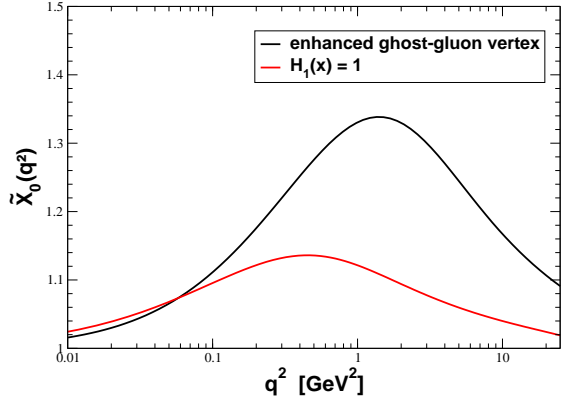} 
   \caption{The theoretical dressed one-loop evaluation of the form factor $\xoeff (p^2)$; see text for details.}
   \label{fig:x0-oneloopdressed}
\end{figure}

The solutions for $\xoeff$ about to be discussed assume a particular value for $\xoeff ( \Lambda^2 )$. Since we are limited by the available momentum 
range of the lattice propagators, we resort to theoretical constraints which we impose on $\xoeff (q^2)$ in the UV. In order to make a connection with the 
perturbative result, we use the one-loop dressed approximation for the quark-ghost scattering kernel, see Fig.~\ref{fig:quark_ghost}, and improve 
the approximation discussed in Ref.~\cite{Aguilar:2010cn}. Indeed, $H(p_1,p_2,p_3)$ is computed solving self-consistently the diagram in Fig.~\ref{fig:quark_ghost} 
for $X_0(p^2)$ and replacing the tree level gluon-ghost vertex by a dressed vertex~\cite{Dudal:2012zx}, which is  parametrized by the form factor,
\be
 H_1 (x) =  c \left( 1 + \frac{a^2 x^2}{x^4 + b^4} \right) + (1 - c)\, \frac{w^4}{w^4 + x^4} \ ,
 \label{Eq:H1}
\ee
with $ c = 1.26$, $a = 0.80$~GeV, $b = 1.3$~GeV and $w = 0.65$~GeV. Therefore, $\xoeff (p^2)$ is given by the solution of the following integral equation:
\bee
 X_0 (p^2) & = &  {\textstyle \frac{1}{4}} \ \mathrm{Tr_{CD}} \ H(-p/2, -p/2, p) 
                 =   1 + \frac{C_F g^2}{8} \int_k^\Lambda \left[p^2 - \frac{(k \cdot p)^2}{k^2}�\right]   \frac{\Delta (k) F(k+p/2) \, F(k)}{(k + p/2)^2} \nonumber \\
         & \times &  \frac{A(k+p) \, \left[ A(k+p) + A(p) \right] H_1 \left( (k+p)^2 \right)\! X_0 \left( k^2 \right)}{A^2(k+p) \, (k+p)^2 + B^2(k+p)}  \, .
  \label{Eq:X0_final}
\eee
The replacement of the tree-level ghost-gluon vertex by a dressed vertex enhances $\xoeff$ by about 20\% and is momentum dependent;
see Fig. \ref{fig:x0-oneloopdressed}.  However, even so this enhancement does not produce sufficient DCSB for a constituent quark mass of the 
expected order of magnitude,  i.e.~$M(0) \approx 300$~MeV.

\subsubsection{Inversion solutions for $\mathbf{\xoeff}$ \label{SubSubSec:LinearReg}}

The above procedure prescribes a normalization for $\xoeff$ at the cutoff which is $\xoeff( \Lambda^2 ) = 1.030$ for $\Lambda = 5$~GeV,  
$\xoeff ( \Lambda^2 ) = 1.024$ for $\Lambda = 6$~GeV and $\xoeff ( \Lambda^2 ) = 1.021$ for $\Lambda = 7$~GeV.

\begin{figure}[t] 
\vspace*{-1cm}
   \centering 
   \includegraphics[scale=0.7]{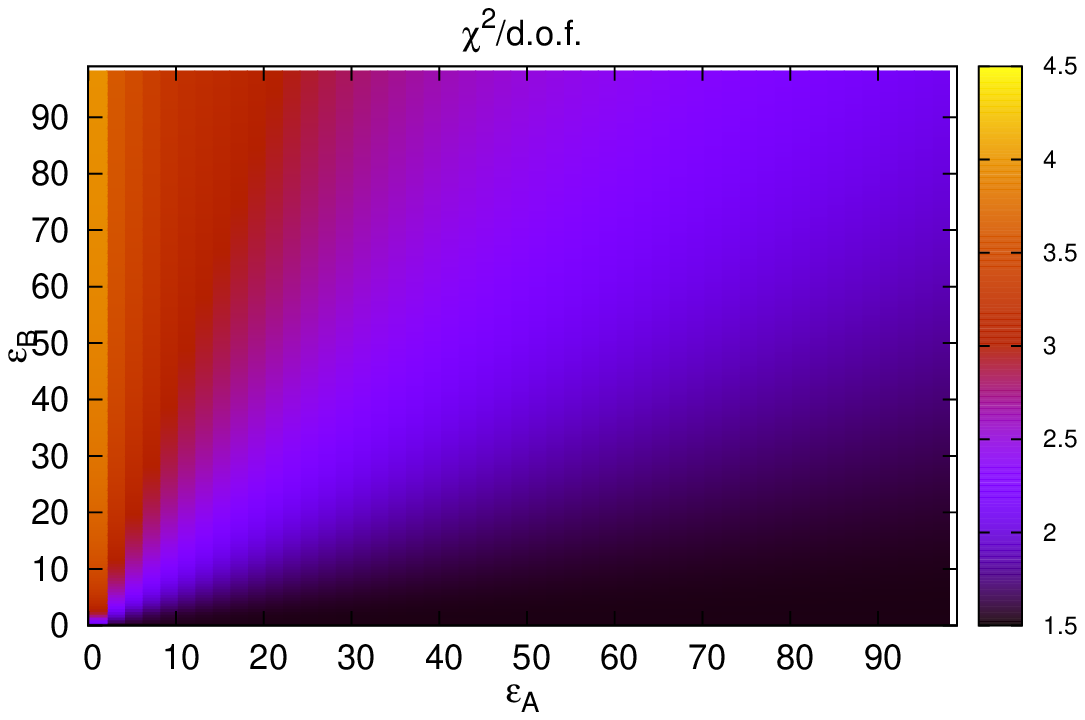} \\  \vspace*{-7mm}
   \includegraphics[scale=0.7]{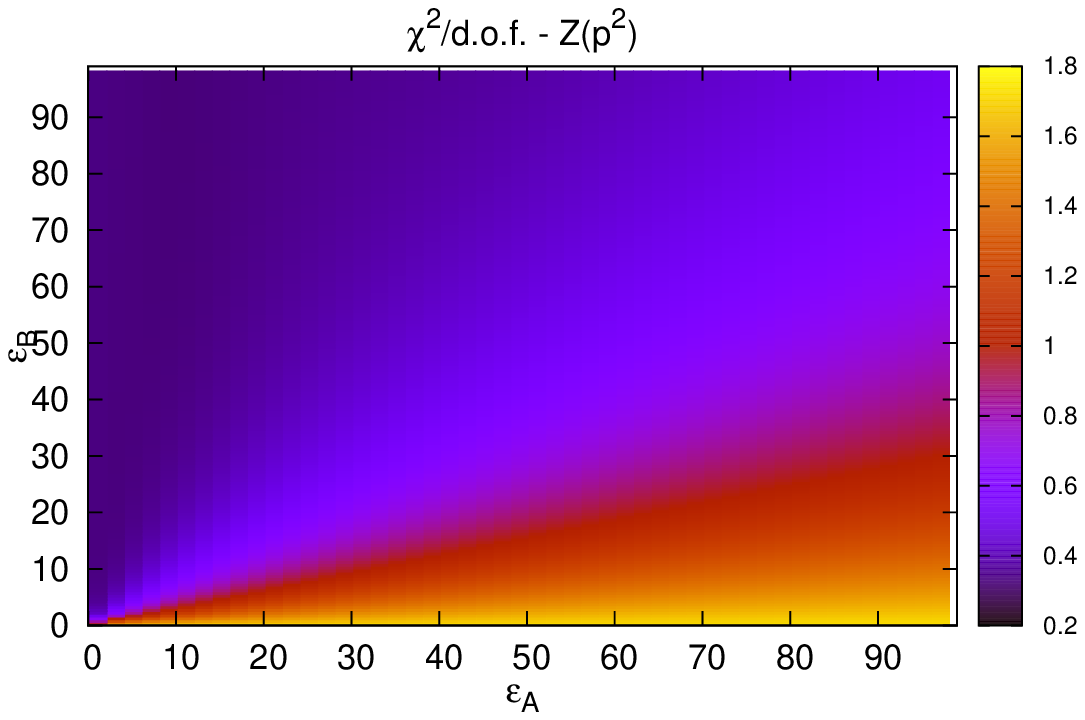} \\  \vspace*{-7mm}
   \includegraphics[scale=0.7]{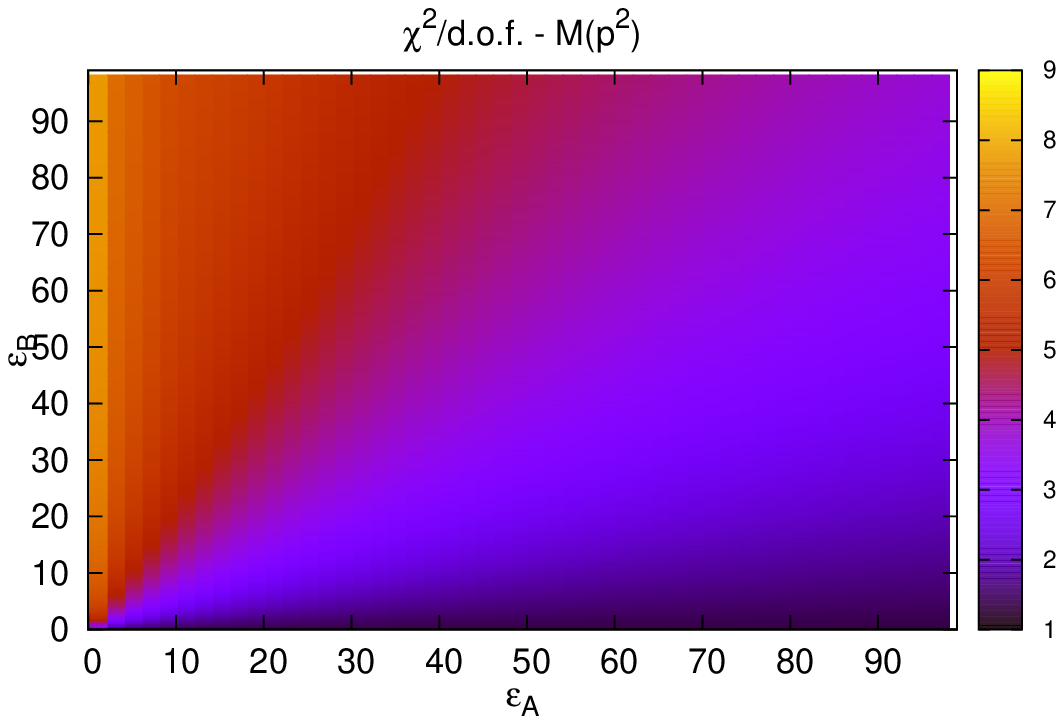}  \vspace*{-3mm}
   \caption{The $\epsilon_A/\epsilon_B$ plane of $\chi^2/\dof$ values obtained in the inversion process of the DSEs. The two lower graphs depict 
   the results for $Z(p^2)$ and $M(p^2)$, whereas the upper graph describes the summed $\chi^2/\dof$, see Eqs.~(\ref{chi2I}) to (\ref{chi2II}). 
   All results are for $\Lambda = 5$~GeV. }
   \label{fig:inversion_chi2}
 \vspace*{-3mm}  
 \end{figure}

As mentioned previously, we judge the quality of the $\xoeff (q^2)$ solution via a $\chi^2$-comparison of $A(p^2)$ and $B(p^2)$ (in whose expressions 
$\xoeff (q^2)$ enters) and therefore $Z(p^2) = 1/A(p^2)$ and $M = B(p^2)/A(p^2)$ with the corresponding lattice values.
We thus evaluate the following $\chi^2/\dof$:
\bee
  \chi^2_Z    & = & \sum_i \frac{\left |  Z(p^2_i) - Z_{\mathrm{\,Latt.}}(p^2_i)\right |^2 }{\sigma^2_Z(p^2_i)}   \ , \label{chi2I}  \\
  \chi^2_{M} & = & \sum_i \frac{\left | M(p^2_i) - M_{\mathrm{\,Latt.}}(p^2_i)\right |^2}{\sigma^2_M(p^2_i)}  \ , \\
  \chi^2  & = & \chi^2_Z + \chi^2_M \ . \label{Eq:chi2comb} \label{chi2II} 
\eee
The momenta $p^2_i$ are those corresponding to the lattice data and $\sigma_{Z,M} (p_i^2)$ denotes the statistical errors. The above quantities are normalized 
to the number of data points. The various $\chi^2/\dof$ values for $\Lambda = 5$~GeV as a function of the regularization parameters 
$\epsilon_A$ and $\epsilon_B$ are shown in Fig.~\ref{fig:inversion_chi2}.

The quark wave function is relatively insensitive to the values chosen for $\epsilon_A$ and $\epsilon_B$. Indeed, changing both regularization parameters 
by two orders of magnitude results in acceptable values for $\chi^2_Z/\dof$ The inversions with $\chi^2_Z/\dof \approx 1$ always feature $\epsilon_A > \epsilon_B$. 
On the other hand, the minimizatin of $\chi^2_M/\dof$ clearly prefers $\epsilon_A \gg \epsilon_B$  and favors smaller $\epsilon_B$ values. A too large regularization 
parameter  $\epsilon_B$ in the $B(p^2)$ equation significantly increases $\chi^2_M/\dof$, in other words the confidence level of the ``fit" is low.  
The upper plot in Fig.~\ref{fig:inversion_chi2} shows the combined $\chi^2/\dof$ as defined in Eq.~(\ref{Eq:chi2comb}). Good solutions, in the sense of 
having $\chi^2/\dof \approx 1$, point towards small  $\epsilon_B$ values and are less sensitive to the choice of $\epsilon_A$.

\begin{figure}[t!] 
\vspace*{-6mm}
   \centering
   \includegraphics[scale=0.32]{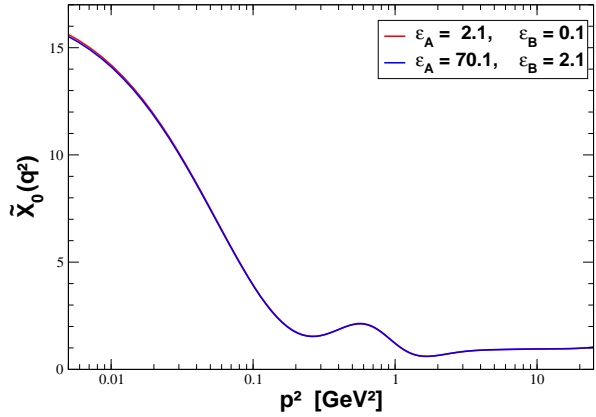} 
   \caption{The form factor $\xoeff$ from inverting the DSE using two different linear regularizations. See text for details.}
   \label{fig:X0_linear_regression} \vspace*{5mm}
\end{figure}

\begin{figure}[h] 
  \centering
  \hspace*{-7mm}
   \includegraphics[scale=0.3]{Z_Lambda5GeV_min_eps_small_eps.eps} \\ \vspace*{-6mm}
   \includegraphics[scale=0.3,angle=-90]{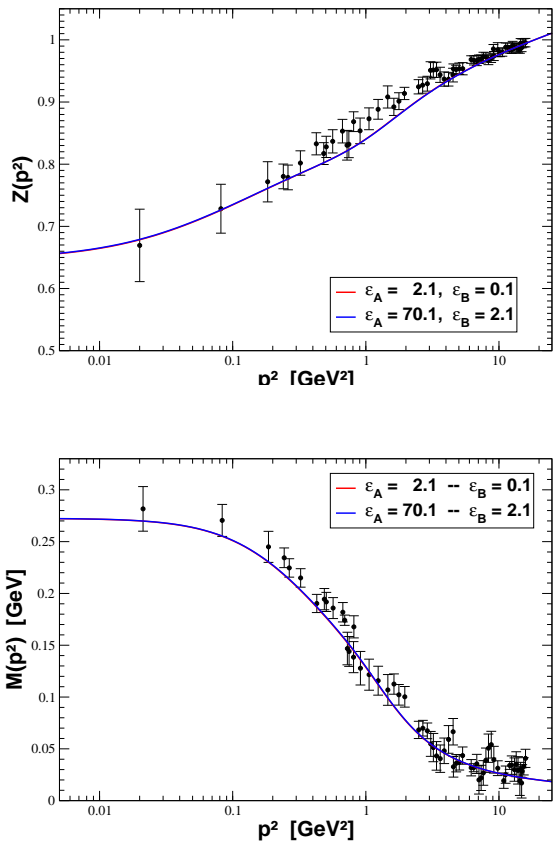} 
   \caption{The quark wave function (upper plot) and quark mass function (lower plot) calculated with $\xoeff (p^2)$ 
     (see Fig.~\ref{fig:X0_linear_regression}) and the r.h.s of the DSE, Eqs.~(\ref{EQ:SDE_B}) and (\ref{EQ:SDE_A}).}
   \label{fig:Z_and_M_linear_regression}
\end{figure}

For $\Lambda = 5$~GeV, we consider two different solutions: {\em i\/}) the solution with minimal $\chi^2/\dof$, which occurs for $\epsilon_A = 70.1$,  
$\epsilon_B = 2.1$ and whose quality is expressed by $\chi^2/\dof= 1.536$, $\chi^2_Z/\dof =  1.642$, $\chi^2_M/\dof =  1.427$; {\em ii\/}) the solution 
with the smallest regularization parameters but similar $\chi^2/\dof$, i.e. $\epsilon_A =  2.1$, $\epsilon_B =   0.1$ and $\chi^2/\dof  =  1.537$, 
$\chi^2_Z/\dof =  1.587$, $\chi^2_M/\dof = 1.486$. The form factor $\xoeff$ for this two sets 
are plotted in Fig.~\ref{fig:X0_linear_regression}. The corresponding quark wave function and running quark mass are given in Fig.~\ref{fig:Z_and_M_linear_regression}. 
As these figures show, the results of the inversion are stable against variation of $\epsilon_A$ and $\epsilon_B$ and the curves are indistinguishable. 
However, in the deep IR domain, not depicted in Figs.~\ref{fig:X0_linear_regression} and \ref{fig:Z_and_M_linear_regression}, one observes a slight difference 
between the solutions.

The form factor $\xoeff (p^2)$ is a monotonous function with some structure around $p^2 \simeq 0.6$~GeV$^2$, reaching a value of $\xoeff (p^2) \approx 2$ 
and increasing continuously from $p^2 \simeq 0.15$~GeV$^2$ as one approaches zero momentum. The solution does not grow indefinitely in the IR and at 
$p^2 \approx 10^{-3}$~GeV$^2$ it reaches a plateau where $\xoeff \simeq 17$.  We do not plot $\xoeff$ in this deep IR region where lattice data do not 
provide any information and whose contribution to the DSE is negligible. For momenta above $p^2 \simeq 1.1$~GeV$^2$, $\xoeff$ slightly increases and reaches 
its  ``perturbative" value at the cutoff, $\Lambda$.

The functional form of $\xoeff (p^2)$ in Fig.~\ref{fig:X0_linear_regression} is rather different from the form factor $X_0(q^2)$ computed using a one-loop dressed 
approximation; see Fig.~6 in Ref.~\cite{Aguilar:2010cn}. The latter form factor barely deviates from $X_0(p^2) =1$ and takes on a single 
local maximum, $X_0(p^2) \simeq 1.25$ for $p^2 \simeq 0.12$ GeV$^2$. At larger momenta, $X_0(p^2)$ recovers the perturbative value 
and approaches one; similarly, the form factor reaches unity in the deep IR. Our form factors show a considerably richer structure, 
with a local maximum at $p^2 \simeq 0.5$~GeV$^2$ where $\xoeff \simeq 2$, a local minimum at $p^2 \simeq 1.1$~GeV$^2$ ($X_0 \simeq 0.5$) 
and another minimum at $p^2 = 0.12$~GeV$^2$ ($X_0 \simeq 1.5$). An absolute maximum is located in the deep IR region, where  we find 
$\xoeff \simeq 17$.  We would like to recall the reader that in order to generate the necessary DCSB in their DSE for the quark, the authors 
of Ref.~\cite{Aguilar:2010cn} had to replace $Z_c^{-1}\mathcal{K\,}_0 (q^2) \to F(q^2)\mathcal{K\,}_0(q^2)$ (or equivalenlty $F(q^2) \to F^2(q^2)$) in the kernel
to enhance its IR strength by as much as a factor of $\sim 3$. The form factor $\xoeff (q^2)$ extracted from the inversion not only has a much 
richer structure but also enhances the quark-gluon interaction compared to the dressed one-loop approximation.

\begin{figure}[t] 
\vspace*{-6mm}
   \centering
   \includegraphics[scale=0.32]{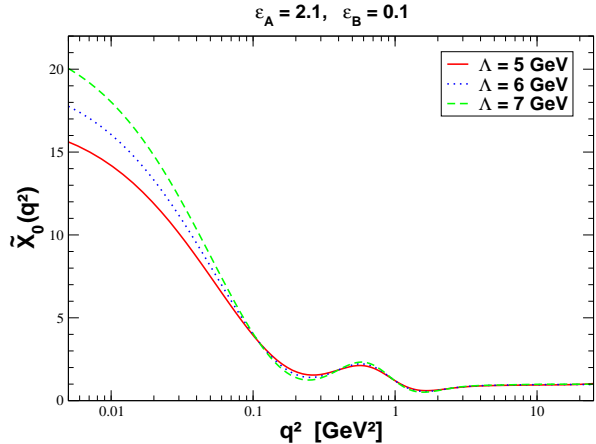}
   \caption{The effective form factor $\xoeff (q^2)$ for three values of the cutoff $\Lambda$.}
   \label{fig:X0_various_lambda}
\end{figure}

\begin{figure}[t]
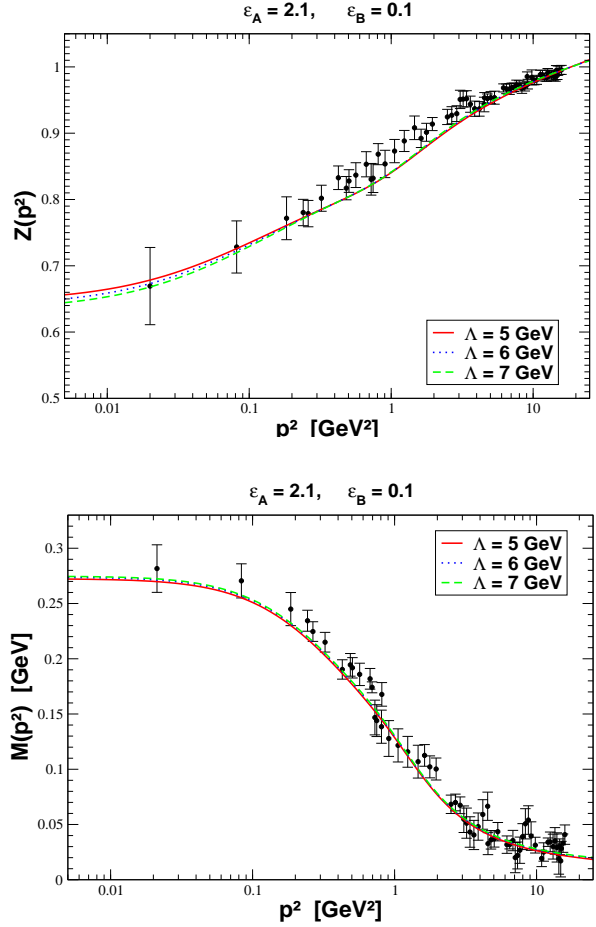
 
\vspace*{-5mm}
   \centering
   \includegraphics[scale=0.32]{Z_various_Lambda.eps}    \\  \vspace*{5mm}
    \includegraphics[scale=0.32]{M_various_Lambda.eps}
   \caption{The quark wave function (upper plot) and the mass function (lower plot) for various cutoff values.}
   \label{fig:Z_M_various_lambda}
\end{figure}

In Fig.~\ref{fig:Z_and_M_linear_regression}, we compare the quark wave function and mass function obtained with $\xoeff (q^2)$ and the DSE. 
More precisely, we make use of $\xoeff (q^2)$, $A_\mathrm{Latt.}(p^2)$ and $B_\mathrm{Latt.}(p^2)$ and compute the integrals of the r.h.s of 
Eqs.~(\ref{EQ:SDE_B}) and (\ref{EQ:SDE_A}) to evaluate $Z(p^2)$ and $M(p^2)$. On the overall, the DSE predictions follow the lattice data with s
mall deviations in $Z(p^2)$ for momenta just above $\simeq 1$~GeV and similarly in $M(p^2)$ for $p^2 \simeq 0.2$~GeV$^2$.
Finally, in Figs.~\ref{fig:X0_various_lambda} and \ref{fig:Z_M_various_lambda}, we investigate how $\xoeff (p^2)$, $Z(p^2)$ and $M(p^2)$ depend 
on the cutoff, $\Lambda$, for $\epsilon_A = 2.1$ and $\epsilon_B = 0.1$. The $\xoeff$ solutions are characterized by $\chi^2/\dof = 1.357$, 
$\chi^2_Z/\dof = 1.388$, $\chi^2_M/\dof = 1.325$ for $\Lambda = 6$~GeV and $\chi^2/\dof = 1.259$, $\chi^2_Z/\dof = 1.292$, $\chi^2_M/\dof = 1.225$ 
for $\Lambda = 7$~GeV.

As can be read from Figs.~\ref{fig:X0_various_lambda} and \ref{fig:Z_M_various_lambda}, the inversion is independent of the cutoff for momenta 
within the range of available lattice data for $Z(p^2)$ and $M(p^2)$. The variation of $\xoeff$ with $\Lambda$, relative to the mean value of 
the various estimates, is about 11\% for $p^2 \lesssim 0.115$~GeV$^2$, whereas for larger momenta it is below the 3\% level. We remind that the 
lattice-regularized propagators cover a relative short momentum range and extending the lattice fits to larger momenta,  as required for larger 
cutoffs can introduce some bias on the results. Furthermore, as discussed in Section~\ref{Sec:kernels_dse}, the IR region is much less constrained 
in the inversion process, therefore $\xoeff$ can differ considerably in this region without a significant change in the quark wave and mass function; 
this is clearly observed in Fig.~\ref{fig:Z_M_various_lambda}. Nonetheless, our numerical checks demonstrate that, for a fixed cutoff, the inversion 
of the DSE for $\xoeff$ is stable and independent of the regulators of the linear system, namely $\epsilon_A$ and $\epsilon_B$. 
   
\subsection{$\mathbf{\xoeff}$ from the Maximum Entropy Method \label{SubSec:MEM}}

Another approach to determine $\xoeff (q^2)$ is to express the form factor in a given functional basis. We do so 
by writing $\xoeff (q^2)$ as a superposition of rectangular-like functions, $r_m(q^2)$,
\bee
   \xoeff (q^2)=\sum_{m=1}^{M=47} x_m r_m(q^2)  \ ,
\label{EQ:x0}
\eee
where for $k=3$~GeV$^{-1}$,
\bee
  r_m(q^2) =\frac{1}{\Big [1+e^{-2k(q-q^\mathrm{min}_m) }  \Big ]  \Big [ 1+e^{-2k(q^\mathrm{max}_m-q) }  \Big ] }  \  ,
\eee
and $q^\mathrm{max}_m = (q_{m+1}+q_m)/2$,  $q^\mathrm{min}_m =  (q_{m-1}+q_m)/2$. The sequence, $q_1 < q_2 < \cdots < q_M$,
is a  momentum partition of the integrand and $x_m$ are free parameters. In the limit $k \rightarrow \infty$, the function $r_m(q)$ vanishes 
over the entire real line except in the intervals, $q \in \left[ q^\mathrm{min}_m,q^\mathrm{max}_m \right]$, where it is unity. Inserting the expansion 
of Eq.~(\ref{EQ:x0}) into Eqs.~(\ref{EQ:SDE_B}) and (\ref{EQ:SDE_A}), the angular-momentum integration can be performed and we are left 
with $A(p^2)$ and $B(p^2)$ expressed as a linear superposition of $x_m$, 
\bee
A (p^2) & = &  Z_2 + \sum_m\, x_m\, A_m (p^2)  \ , \\
B (p^2) & = &  Z_4 \,  m (\mu) + \sum_m\,  x_m\,  B_m (p^2) 
\label{BMEM}
\eee
where $B_m$ and $A_m$ merely depend on the external momenta defined by the lattice simulation. The coefficients $x_m$ can be determined 
via a $\chi^2$-fit to the lattice data points, $Z_\mathrm{Latt.} (p^2_i)$ and $M_\mathrm{Latt.} (p^2_i)$, i.e.
\begin{equation}
 \chi^2_{\scriptscriptstyle{A,B}}  =
            \sum_i   \frac{ \left | Z_{\mathrm{Latt.}} (p^2_i) -  Z(p^2_i,r_m,x_m)\right |^2 }{\sigma_{\scriptscriptstyle{Z}} (p^2_i)}   
              +  
            \sum_i \frac{ \left | M_{\mathrm{Latt.}} (p^2_i) -  M(p^2_i,r_m,x_m) \right |^2}{\sigma_{\scriptscriptstyle{M}} ( p^2_i)} \,  ,
\end{equation}
where a sum over $m$ is implicit in the expressions for $Z(p^2_i,r_m,x_m)$ and $M(p^2_i,r_m,x_m)$.
The statistical errors on $Z_{\mathrm{Latt.}}$ and $M_{\mathrm{Latt.}}$ are given by $\sigma_{\scriptscriptstyle{Z,M}} (p_i^2)$.
With the renormalization condition in Eq.~(\ref{renormcondition}), we obtain $Z_4=0.71$ which corresponds to
$Z_4 (\mu,\Lambda )\, m(\mu)  =  Z_2 (\mu,\Lambda )\, m^{\rm bm} (\Lambda) = 17.7$ MeV.

As discussed in Section~\ref{Sec:kernels_dse}, the DSE kernels  are not invertible and to define a solution one must regularize the corresponding linear system. 
In this section, we choose an alternative route and consider the Maximum Entropy Method (MEM). With MEM, one introduces the {\em negative entropy\/} of
 the solution,
\be
  \lambda \sum_m\,  \xoeff \left (r_m(q^2) \right )\, \log \xoeff \left (r_m(q^2) \right ) \ ,
\ee
to be added to the $\chi^2$ function. We also find useful to add the term,
\be 
  \lambda \sum_m\,  x_m\, \log x_m \ ,
\ee  
owing to the fact that the $x_m$ are the minimization parameters.  The parameter $\lambda$ is adjusted, 
\begin{eqnarray}
  \chi^2 ( \lambda ) & = & \chi^2_{\scriptscriptstyle{A,B}} +
             \lambda \, \sum_m \, \Big \{ \,  \xoeff \left (r_m(q^2) \right )\, \log \xoeff \left (r_m(q^2) \right )  +  x_m\, \log x_m \Big\} \ ,
\end{eqnarray}
so that  $\chi^2_{A,B} \sim N$, where $N$ is the number of lattice-data points, which yields a smooth solution. 
Note the assumption is that $\xoeff$ and $x_m$ are  positive definite.

\begin{figure}[t] 
   \centering  
   \vspace{0.6cm} 
   \includegraphics[scale=0.34]{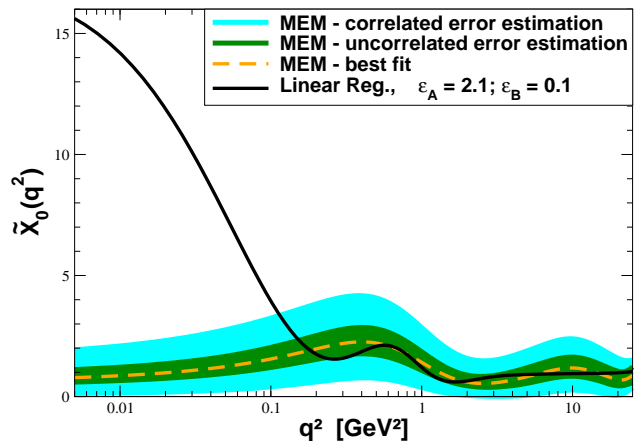} 
   \caption{The effective form factors $\xoeff$ computed with the Maximum Entropy Method. The dashed curve refers to best fit to 
               the enlarged data set.  The green error band describes the root-mean-squared deviation, whereas the cyan-shaded area is the maximum 
               uncertainty for $\xoeff$ assuming full correlation, see Eq.~(\ref{rms}) and (\ref{maxerr}).
            The prediction of the linear regularization method (see Section~\ref{DSEregularized}) is given by the black solid curve.}
   \label{fig:x0mem}
\end{figure}

From the lattice-quark propagator~\cite{Bowman:2005vx}, we extract a total of 124 data points: 61 points for the mass function and 63 points for the quark wave function. 
The data set is enlarged by taking 7 momenta above $4$~GeV exploiting the fit to the lattice data (see Section~\ref{dynquarkprop}) up to 5~GeV. The 47 parameters,
$x_m$, are fitted to $N=138$ data points while $\lambda$ is adjusted, which yields $\chi^2 = 139.4$. We remind that within MEM the adjustment of $\lambda$ does
not necessarily provide the optimal $\chi^2/\dof$ Indeed, whilest $\chi^2/\dof \simeq 1$ can be achieved in a best fit, there are numerous corresponding solutions 
for $\xoeff$ owing to the analytic structure of the kernels, see Eqs.~(\ref{EQ:kernel_A}) and (\ref{EQ:kernel_B}) and Fig.~\ref{fig:kernels}.

The $\xoeff$ for the best fit to the enlarged data set is depicted by the dashed curve in Fig.~\ref{fig:x0mem}. Hereafter, we refer to this solution 
as the MEM-improved $\xoeff$. If one assumes that the $x_m$ are uncorrelated, i.e. the correlation matrix is given by $\rho = \delta_{mm^\prime}$, 
then the uncertainty in $\xoeff$ is given by,
\bee
\sqrt{\sum_{mm^\prime}\frac{\partial \xoeff}{\partial x_m}\rho_{mm^\prime}  \sigma_m \sigma_{m^\prime} \frac{\partial \xoeff}{\partial x_{m^\prime}} }
   =\sqrt{\sum_m\left( \frac{\partial \xoeff}{\partial x_m}\right )^{\!\!\! 2} \sigma_m^2} \ ,
   \label{rms}
\eee
where $\sigma_m$ is the statistical error on $x_m$. This uncertainty defines the green error band in Fig.~\ref{fig:x0mem}.
On the other hand, if one assumes that all $x_m$ are correlated, i.e. $\rho_{mm^\prime}=1$, the uncertainty in the form factor becomes,
\bee
\sqrt{\sum_{mm^\prime}\frac{\partial \xoeff}{\partial x_m}\rho_{mm^\prime}  \sigma_m \sigma_{m^\prime} \frac{\partial \xoeff}{\partial x_{m^\prime}} }
        =\sum_m \frac{\partial \xoeff}{\partial x_m} \sigma_m \ .
 \label{maxerr}
\eee
which is represented by the cyan-shaded band in Fig.~\ref{fig:x0mem}.

\begin{figure}[t]
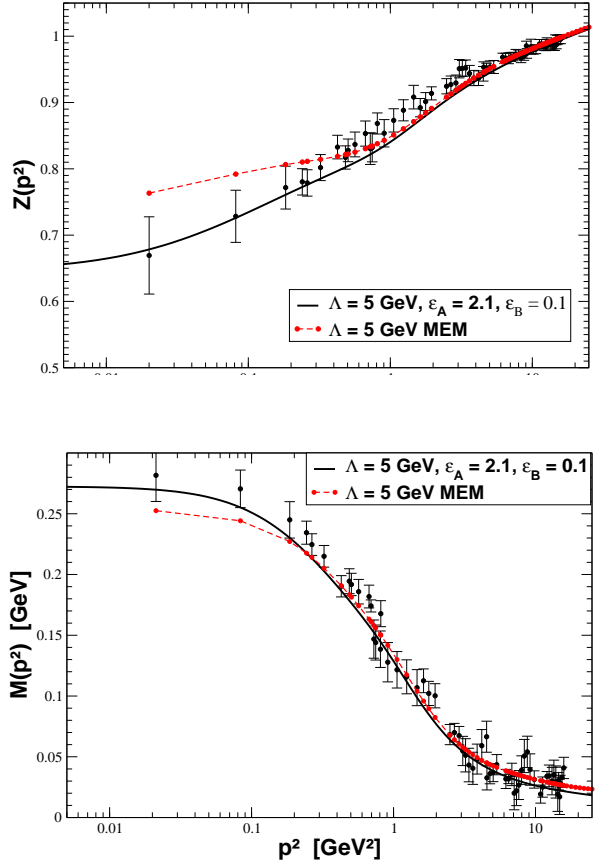
 
  \vspace*{-5mm}
   \centering
   \includegraphics[scale=0.32]{Z_various_Lambda2_nova.eps} \\  \vspace*{5mm}
   \includegraphics[scale=0.32]{M_various_Lambda2_nova.eps} 
   \caption{The quark wave function (upper plot) and mass function (lower plot) computed with the two regularization methods
                 for a cutoff $\Lambda = 5$ GeV.}
  \label{fig:Z_M_various_lambda_nova}
\end{figure}

The two inversion methods discussed are different in nature and provide the same functional form of $\xoeff (q^2)$ for momenta above 
$q^2 \approx 0.1$~GeV$^2$. In the deep IR region, the values for $\xoeff$ obtained with MEM and the chosen basis of Eq.~(\ref{EQ:x0}) lie
considerably below those of the linearly regularized solution. The differences in the IR region are reflected in the quark wave and 
mass function, as becomes clear from Fig.~\ref{fig:Z_M_various_lambda_nova}, where the lattice data is also provided. While the running quark 
mass is less sensitive to the inversion method --- recall that it is given by the ratio $B(p^2) / A(p^2)$ --- the quark wave function clearly 
reveals the infrared structure of the quark-gluon vertex below $q^2 \sim 0.1$~GeV$^2$. Our results suggest that $\xoeff (p^2)$ should 
be enhanced at very small momenta. Note that at the level of one standard deviation, neither method is able to reproduce the lattice for 
$Z(p^2)$ for $p^2$ in the range $1 - 4$ GeV$^2$. The missing strength of the MEM solution in the low momenta region of $\xoeff$ can 
also be observed in the running quark mass compared to the linear regularized solution and the lattice data.



\section{Closing the gap: solving the DSE with $\mathbf{\xoeff}$ \label{Sec:Closing}}

Let us now discuss the self-consistent solution of Eqs.~(\ref{EQ:SDE_B}) and (\ref{EQ:SDE_A}) for the dressed-quark propagator, i.e. for 
the functions $A(p^2)$ and $B(p^2)$, using the linear regularized solution (LRS) for $\xoeff$ obtained in Section~\ref{SubSubSec:LinearReg}
and the MEM solution of Section \ref{SubSec:MEM}. We remind that in both cases the lattice gluon and ghost propagators, described in 
Section~\ref{SubSec:gluonghostprops}, are inputs for the DSE.

The self-consistent quark functions $A(p^2)$ and $B(p^2)$ are computed imposing the renormalization conditions
$B(\mu^2)= B_\mathrm{Latt.}(\mu^2)$ and $A(\mu^2)= A_\mathrm{Latt.}(\mu^2)=1$. Consequently, the following renormalization constants, 
$Z_2$ and $Z_4$, are found:
\begin{equation}
Z_4 = \left\{ \begin{array}{l@{\hspace{0.5cm}}l} 0.75 & \mbox{LRS} \\ 0.72 & \mbox{MEM} \end{array} \right.
\qquad\qquad\mbox{ and }Ê\qquad\qquad
Z_2 = \left\{ \begin{array}{l@{\hspace{0.5cm}}l} 0.938 & \mbox{LRS} \\ 0.936 & \mbox{MEM} \end{array} \right. \ .
\end{equation}

\begin{figure}[t]
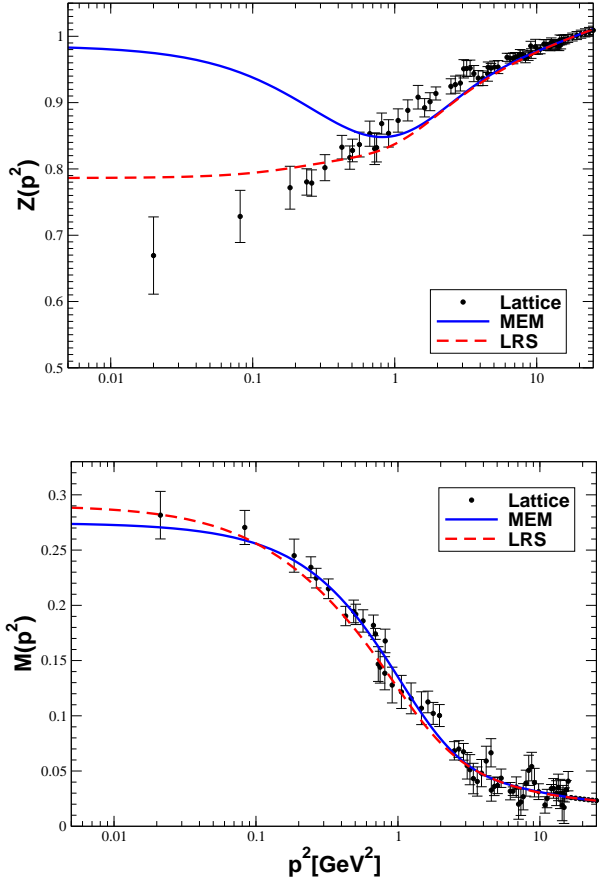
 
  \vspace*{-5mm}
   \centering
   \includegraphics[scale=0.32]{Z-plot_new_SDE.eps}   \\  \vspace*{5mm}
   \includegraphics[scale=0.32]{M-plot_new_SDE.eps} 
   \caption{ The quark wave function (upper plot) and mass function (lower plot) from solving the DSE, Eqs.~(\ref{EQ:SDE_B}) and Eq.~(\ref{EQ:SDE_A}), 
                 with $\xoeff (p^2)$ from both the LRS and MEM.  The cutoff is $\Lambda = 5$~GeV and the lattice data is taken from Ref.~\cite{Bowman:2005vx}. }
   \label{fig:Z_M_from_SDE_newplots}
\end{figure}

The DSE solutions are show in Fig.~\ref{fig:Z_M_from_SDE_newplots} for the kernels computed with $\xoeff$ from LRS and MEM. The mass function
provided by the two solutions is of about the same quality. However, the quark wave function shows large deviations with respect to the 
lattice data in the IR. The difference between the two quark functions, $Z(p^2)$ and $M(p^2)$, can be understood as follows: altering $A(p^2)$ and 
$B(p^2)$ simultaneously and similarly in the IR, the ratio $M(p^2) = B(p^2)/A(p^2)$ remains less sensitive to the approximations in our theoretical 
framework. Concerning the quark wave function, it is well known that this quantity is rather sensitive to the different DSE truncations in the IR --- see, 
e.g., Fig.~4.13 in Ref.~\cite{williamsphd}. Our solutions of the DSE are also in qualitative agreement with those of Ref.~\cite{Aguilar:2010cn}.

The difference in $Z(p^2)$ for $p^2 \lesssim 0.7$~GeV$^2$, reported in Fig.~\ref{fig:Z_M_from_SDE_newplots}, is due to the 
distinct behavior of the form factors $\xoeff (q^2)$ of the two methods, which is maximal in the low momenta region. 
The LRS enhances significantly the infrared quark-gluon vertex producing a quark wave function closer to the lattice points. 

The departure from the lattice data of the DSE solutions is a direct consequence of all approximations and assumptions employed, namely the
non-Abelian form of the Ball-Chiu vertex and kinematic simplifications in the definition of $\xoeff (q^2)$. More precisely, the calculation does 
not take into account the transverse components of the quark-gluon vertex, the form factors $X_1$, $X_2$, $X_3$ and the full kinematical 
dependence of $X_0 (p_1,p_2,p_3) \approx X_0(q^2)$, where $q=p_3$ is the gluon momentum in the quark-gluon vertex.  In this sense, 
it is surprising that the approximation is able to qualitatively reproduce the lattice results.

%

\section{Summary \label{conclusion}}

In this work, we investigated the quark-gluon vertex in the IR region combining two non-perturbative QCD techniques:
DSEs are employed along with lattice simulations to extract information on this fundamental QCD vertex.

The general structure of the dressed quark-gluon vertex includes twelve independent form factors. We use an ansatz compatible with the 
STI of Eq.~(\ref{STI}) which does not take into account the transverse components of the vertex. This ansatz has a generalized Ball-Chiu form, 
parameterized in terms of four forms factors, $X_i$, and the ghost dressing function, $F(p^2)$. The $X_i$ are associated with the tensor 
structure of the quark-ghost scattering kernel. The model assumes that $X_0$, the dominant form factor in the perturbative solution of QCD, 
provides the main contribution to the vertex and the drastic approximations, $X_0 \approx X_0 (q^2)$ and $X_1 = X_2 = X_3 = 0$ are made, 
where $q$ is the gluon four-momentum of the quark-gluon vertex.

We considered the DSE using the lattice dressed-quark, -gluon and -ghost propagators as inputs to extract $X_0 (q^2)$ which includes 
contributions from $X_i$ , $i = 1$, 2, 3, as well as from the transverse components of the vertex. 
Our result should be interpreted as an effective form factor, $\xoeff (q^2)$, rather than the $X_0$ of the quark-ghost scattering kernel.
The kernels of the integral equations that contain $\xoeff$ are not invertible. Indeed, they have small or vanishing eigenvalues. This problem 
is overcome by regularizing the DSE, now transformed into a linear system of equations. In order to extract $\xoeff$, we applied two distinct 
and independent methods. Our first approach to the inversion of the gap equation is expressed via a linear regularized equation 
solved by iteration. The second approach is based on a decomposition of $\xoeff$ in terms of a functional basis whose 
parameters, $x_m$, are adjusted to reproduce the lattice-regularized QCD results for $M(p^2)$ and $Z(p^2)$ via a $\chi^2$-fit 
combined with the Maximum Entropy Method. 

The two methods produce $\xoeff$ form factors compatible with each other for the range of momenta where lattice-simulation data is 
available, i.e. in the domain $\simeq 0.3-4$~GeV.  For momenta in the range $200$~MeV$-1$~GeV, both regularizations feature a strong 
enhancement of the generalized Ball-Chiu vertex, as expected in order to generate the DCSB observed in the lattice mass function.
However, for momenta below $\sim 300$ MeV, the $\xoeff$ obtained with the linear regularization is characterized by a much stronger 
enhancement in comparison with the MEM result which essentially reproduces the  Ball-Chiu vertex, i.e. $\xoeff \approx 1$ at such low 
momenta.  For large enough momenta, both methods recover the perturbative value, $\xoeff \approx 1$.

The respective $\xoeff (q^2)$ are inserted in the r.h.s. of the DSE, along with the lattice quark, gluon and ghost propagators, to numerically 
evaluate the integrals for $Z(p^2)$ and $M(p^2)$. The latter are then compared to the corresponding lattice data. It turns out that the 
quark mass function does not distinguish between the two solutions, while the wave function favors the strong enhancement of $\xoeff$ 
found,  for very small momenta, in the linear regularized solution of the DSE.

The self-consistent solution of the quark DSE, using a kernel built upon the generalized Ball-Chiu vertex with $\xoeff (p^2)$ and the lattice 
gluon and ghost propagators, provides the required DCSB to adequately describe the lattice mass function of Ref.~\cite{Bowman:2005vx}. 
However, it fails to reproduce the lattice quark wave function for momenta below $\sim 700$~MeV. We interpret this deviation as due to 
(i) the lack of any significant constraint from the lattice data in the deep IR; (ii) the negligible contribution of the deep infrared momenta 
to the quark gap equation; and, most importantly, (iii) the approximations and assumptions employed in solving the integral 
equations (lack of transverse components, $X_i \approx 0$ for $i = 1$, 2, 3 and kinematic simplifications). In what concerns the 
contributions of the infrared momenta to the gap equation, recall that for practical purposes one may ignore contributions from 
$\xoeff (q^2)$ for $q^2 \lesssim 0.1$~GeV$^2$ in the DSE and still obtain the same solutions of Eqs.~(\ref{EQ:SDE_B}) and (\ref{EQ:SDE_A}). 
Contemporary hadron physics~\cite{Bashir:2012fs,Aznauryan:2012ba} also instructs us that the contribution from the deep IR  domain
in the quark DSE kernel has little impact on relevant hadron observables, such as meson and nucleon electromagnetic form factors.

In short, we have made use of two independent methods which correlate $\xoeff (q^2)$ with the dressed-quark functions, $M_\mathrm{Latt.}(p^2)$ and 
$Z_\mathrm{Latt.}(p^2)$, via a DSE kernel built upon the quark-gluon vertex of Eqs.~(\ref{bcl1})--(\ref{bcl4}) and the quenched gluon and ghost propagators 
from lattice-QCD simulations. The form factor $\xoeff$ enhances the quark-gluon vertex in the IR region and yields $M(0) \simeq 280$~MeV. This is
in contrast with the standard Ball-Chiu ansatz which is equivalent to setting $\xoeff = 1$ and does not produce sufficient DCSB if one employs
a lattice-regularized dressed gluon propagator. Our functional expression for $\xoeff (q^2)$ along with the generalized Ball-Chiu vertex provide an effective 
work tool for quark DSE applications in hadron physics.  We postpone to a future work the full and consistent calculation of all form factors of the 
quark-ghost scattering kernel, $H(p_1,p_2,p_3)$, augmented by the minimal {\em ansatz\/} for the transverse vertex recently put forward by Qin 
et al.~\cite{Qin:2013mta}.

\section*{Acknowledgements}

The authors are supported by the Brazilian agencies FAPESP (Funda\c c\~ao de Amparo \`a Pesquisa do Estado de
S\~ao Paulo) and CNPq (Conselho Nacional de Desenvolvimento Cient\'ifico e Tecnol\'ogico). 
O.O. acknowledges financial support from FCT under contract PTDC/\-FIS/100968/2008, developed under the initiative QREN financed 
by the UE/FEDER through the Programme COMPETE -- Programa Operacional Factores de Competitividade. B.E. benefitted from 
instructive discussions with Adnan Bashir, Jos\'e Rodr\'iguez-Quintero, Craig Roberts and Peter Tandy.


\end{document}